 \documentclass[STIX1COL]{WileyNJD-v2}

\articletype{Article Type}%
\newcommand {\bsis} {\left\{ \begin{array} }
\newcommand {\esis} {\end{array}\right.}

\def\bmat#1{\left[\begin{array}{#1}}
\def\emat{\end{array}\right]}

\newcommand{\blista}{\renewcommand{\labelenumi}{(\roman{enumi})} 
\begin{enumerate}}
\newcommand{\elista}{\end{enumerate} \renewcommand{\labelenumi}{\arabic{enumi}.}}

\def\vv#1{{ \rm \bf{#1}}}
\def\Pr{{\rm{Pr}}}

\newcommand{\setW}{{\mathcal{W}}}
\def\Sum#1#2{\sum\limits_{#1}^{#2}}
\def\R{ {\rm \,I\!R} }    
\DeclareMathOperator*{\argmin}{arg\,min}
\def\unif{\text{unif}}
\def\normal{\text{normal}}

\def\unif{\text{unif}}
\newcommand\thetak{\ensuremath{\theta_{\text{kite}}}} 
\newcommand\phik{\ensuremath{\phi_{\text{kite}}}}
\newcommand\psik{\ensuremath{\psi_{\text{kite}}}}

\def\constraint{g}
\def\QED{{\hfill \rule{2mm}{2mm}}}

\newcommand {\T}{^{\top}} 

\def\conv#1#2{ \left( \begin{array}{c} #1 \\ #2 \\
\end{array}\right)} 

\usepackage{xcolor}
\usepackage{siunitx}
\usepackage{epstopdf}
\usepackage{interval}

\usepackage{comment}
\usepackage[size=normalsize]{subcaption}

\usepackage{tikz}
\usetikzlibrary{trees}
\usepackage{verbatim}

\begin{document}

\title{Probabilistic performance validation of deep learning-based robust NMPC controllers}

\author[1]{Benjamin Karg}

\author[2]{Teodoro Alamo}

\author[1]{Sergio Lucia*}

\authormark{B. Karg \textsc{et al}}

\doi{10.1002/rnc.5696. Copyright by the authors 2021. Published Open Access on the International Journal of Robust and Nonlinear Control by John Wiley \& Sons Ltd.}
\address[1]{\orgdiv{Laboratory of Process Automation Systems}, \orgname{Technische Universit\"at Dortmund}, \orgaddress{Emil-Figge-Str. 70, 44227 Dortmund, \country{Germany}}}

\address[2]{\orgdiv{Departamento de Ingenier\'ia de Sistemas y Autom\'atica}, \orgname{Universidad de Sevilla}, \orgaddress{Escuela Superior de Ingenieros, Camino de los Descubrimientos s/n, 41092 Sevilla, \country{Spain}}}

\corres{*Corresponding author Sergio Lucia. \email{sergio.lucia@tu-dortmund.de}}


\abstract[Summary]{
Solving nonlinear model predictive control problems in real time is still an important challenge despite of recent advances in computing hardware, optimization algorithms and tailored implementations. This challenge is even greater when uncertainty is present due to disturbances, unknown parameters or measurement and estimation errors. To enable the application of advanced control schemes to fast systems and on low-cost embedded hardware, we propose to approximate  a robust nonlinear model controller using deep learning and to verify its quality using probabilistic validation techniques. 

We propose a probabilistic validation technique based on finite families, combined with the idea of generalized maximum and constraint backoff to enable statistically valid conclusions related to general performance indicators. The potential of the proposed approach is demonstrated with simulation results of an uncertain nonlinear system.

}



\maketitle


\section{Introduction}

Model predictive control is a popular advanced control technique that can deal with nonlinear systems and constraints while considering general control goals that go beyond conventional set-point tracking tasks. 
Two of the main obstacles that one faces when implementing and designing a nonlinear model predictive controller are the accuracy of the model and the computational complexity needed to solve a non-convex optimization problem online, which often renders its implementation too slow for fast systems or impossible to be deployed on resource-constrained embedded platforms.

Handling uncertainty in the context of model predictive control is the main goal of robust MPC. Traditional min-max approaches 
\cite{campo1987} do not explicitly consider the fact that new information will be available in the future, which leads to over-conservative solutions. Closed-loop robust MPC avoids the problem of conservativeness by optimizing over control policies instead of optimizing over control inputs~\cite{lee1997}, leading however to intractable formulations in the general case.
Most of the recent robust MPC methods focus on achieving a good trade-off between complexity and performance. 
Tube-based approaches \cite{mayne2005} decompose the robust MPC problem into a nominal MPC and an ancillary controller. The ancillary controller makes sure that the real uncertain system stays close to the trajectory planned by the nominal MPC. By tightening the constraints of the nominal MPC, robust constraint satisfaction can be achieved. In the 
simplest version, the complexity of tube-based MPC is the same as that of standard MPC. However, if an increased performance is desired, the 
complexity grows as presented in  \cite{rakovic2011} or \cite{Fleming2015}. Scenario tree-based \cite{scokaert1998,delaPena2005b, bernardini2009} or multi-stage MPC~\cite{lucia2013} represents the evolution 
of the uncertainty using a tree of discrete uncertainty realizations. An improved performance can be often seen in practice \cite{lucia2014e} because the feedback structure is not restricted to be affine, 
as usually done in tube-based MPC and in other robust approaches~\cite{goulart2006}. 
While it is also possible to achieve stability and robust constraint satisfaction guarantees for a multi-stage MPC formulation \cite{delaPena2005b, LuciaPhd, lucia2020stability}, its computational complexity grows exponentially with the dimension of the uncertainty space. 
The presence of uncertainty significantly increases the computational complexity of any NMPC implementation if a non-conservative performance is desired. 

The last decade has witnessed an important progress on hardware, algorithms and tailored implementations that enable the efficient solution and implementation of NMPC controllers based, for example, on code generation tools \cite{houska2011,mattingley2012cvxgen} that
provide efficient implementations of linear and nonlinear MPC on embedded hardware, including low-cost microcontrollers \cite{zometa2013} and high-performance  FPGAs \cite{lucia2018_TII}.

A different possibility to achieve embedded nonlinear model predictive control is the use of approximate explicit nonlinear model predictive 
control \cite{johansen2004,johansen2002multi} based on approximating the multi-parametric nonlinear program using similar ideas as for explicit MPC of linear systems \cite{bemporad2002}. 
We propose in this work to use deep neural networks to approximate a robust multi-stage NMPC control law. The idea of using a neural network as function approximator for an NMPC feedback law  was already proposed by \cite{parisini1995} back in 1995, but only very recently \cite{karg2018efficient,Chen2018} deep neural networks (neural networks with several hidden layers) have 
been proposed as function approximators. The use of deep neural networks is motivated by recent theoretical results that suggest the exponentially better approximation 
capabilities of deep neural networks in comparison to shallow networks \cite{safran2017}. 


Assessing the closed-loop performance of approximate controllers, or any other controller subject to further random disturbances or estimation errors, is particularly challenging in the case of complex nonlinear systems.
The theory of randomized algorithms \cite{Vidyasagar97}, \cite{Tempo13} provides different schemes capable of addressing this issue. For example, statistical learning techniques can be used to design stochastic model predictive controllers with probabilistic guarantees \cite{Grammatico:16}, \cite{Lorenzen:17}, \cite{Mammarella:18:Control:Systems:Technology}. Also, under a convexity assumption, convex scenario approaches \cite{Calafiore06} can be used in the context of chance constrained MPC \cite{fagiano2010}, \cite{deori2017trading}, \cite{Margellos:14}. 
The main limitation of the aforementioned approaches based on statistical learning results \cite{Vidyasagar97}, \cite{Alamo09} and scenario based ones \cite{Calafiore06} is that the number of random scenarios that have to be generated (sample complexity) grows with the dimension of the problem.

Probabilistic validation \cite{TeBaDa:97}, \cite{Alamo:15}, allows one to determine if a given controller satisfies, with a prespecified probability of violation and confidence, the control constraints.  The sample complexity in this case does not depend on the dimension of the problem, but only of the required guaranteed probability of violation and confidence. Examples of probabilistic verification approaches in the context of control of nonlinear uncertain systems can be found, for example, in \cite{Tempo13},\cite{alamir2018feedback} and \cite{Alamir:15}. These techniques have also been used for the probabilistic certification of the behaviour of off-line approximations of nonlinear control laws \cite{hertneck2018learning}, \cite{zhang2019safe}. 

The main contribution of this paper, which extends the results from \cite{karg2019ecc}, is the formulation of general closed-loop performance indicators that are not restricted to binary functions as in~\cite{hertneck2018learning} and can be computed simulating the closed-loop system with any given controller. We also provide sample complexity bounds that do not grow with the size of the problem for the case of a finite family of design parameters and general performance indicators. Our approach allows to discard a finite number of worst-case closed-loop simulations, improving significantly the applicability of the probabilistic validation scheme compared to existing works. The potential of the presented approach is illustrated for a highly nonlinear towing kite system including a real-time capable embedded implementation of an approximate, but probabilistically safe, robust nonlinear model predictive controller on a low-cost microcontroller.

The paper is organized as follows. The closed-loop performance indicators are introduced in Section~\ref{sec:closed-loop_constraints} which are used in a novel probabilistic validation methodology for arbitrary controllers in Section~\ref{sec:validation}. The mathematical framework for the output feedback robust NMPC problem considered in this work is presented in Section~\ref{sec:robust} and the use of deep learning to obtain approximate robust NMPC controllers is summarized in Section~\ref{sec:deep_learning_approx}. The case study is detailed in Section~\ref{sec:kite_model}, the results in Section~\ref{sec:results} and the paper is concluded in Section~\ref{sec:conclusions}.

\clearpage

\section{Closed-loop performance indicators} \label{sec:closed-loop_constraints}

\subsection{System description and constraints}

We are interested in optimally controlling the following class of nonlinear discrete time systems:
\begin{align}
x(k+1) = f(x(k), u(k), d(k)),\label{eq:LTI}\\
y(k) = h(x(k), u(k), d(k)),\label{eq:meas_function}
\end{align}
where $x(k) \in \mathcal{R}^{n_x}$ is the state vector, $u(k) \in \mathcal{R}^{n_u}$ is the control input, and $d(k) \in \mathcal{R}^{n_d}$ is the disturbance vector. In general, not all states can be measured, and a state estimate $\hat x(k)$ should be computed based on the past measurements $y(k)\in \mathcal{R}^{n_y}$. It is assumed that the disturbances $d(k)$ take values, with high probability, from a known set $\mathcal D$.

\begin{assumption}
    The nonlinear discrete-time system~\eqref{eq:LTI} and~\eqref{eq:meas_function} is observable and controllable.
\end{assumption}


The closed-loop trajectory should satisfy general nonlinear input and state constraints defined by
\begin{align}
    \constraint_{l}(x(k), u(k), d(k)) \leq 0, \; l=1,\ldots,n_g,
\end{align}
where $n_{\constraint}$  is the number of constraints.

\subsection{Closed-loop behavior}

The goal of a controller $\kappa: \mathbb{R}^{n_x} \rightarrow \mathbb{R}^{n_u}$ is that the closed-loop trajectory of the uncertain nonlinear system defined by
\begin{align}\label{eq:closed-loop}
    x(k+1) = f(x(k), \kappa(\hat x(k)), d(k)),
\end{align}
obtains a desired performance level, e.g. does not violate the predefined constraints, despite the presence of uncertainty, for any initial state $x(0)$ in the set $\mathcal{X}_0$ of feasible initial conditions, for any admissible initial estimation error $x(0)-\hat{x}(0)$ and for any sequence of uncertainty realizations $\{d(0), d(1), \dots, d(\infty)\}$. 


Determining if a given controller provides admissible closed-loop trajectories, under the presence of nonlinearity and uncertainty, is in general an intractable problem \cite{blondel2000survey}.
Instead, we focus on the use of finite-time closed-loop performance indicators that can be obtained by simulating the closed-loop system. The underlying assumption is that models which can be run a large number of times are available so that statistical guarantees can be obtained.
A closed-loop performance indicator is defined as follows.

\begin{definition}[Closed-loop finite-time performance indicator]\label{def:performance_indicator}
Let $w = \{x(0), \hat x(0), d(0), \dots, d(N_{\text{sim}}-1)\}$ denote the variables that uniquely define the closed loop trajectories $z(w; N_{\text{sim}}, \kappa) = \{x(0),\hat x(0), \kappa(\hat x(0)), d(0), x(1),\kappa(\hat x(1)), d(1), \dots, x(N_{\text{sim}}) \}$ given an initial condition $x(0)$, an initial estimate $\hat x(0)$, a sequence of uncertainty realizations $d(0), \dots, d(N_{\text{sim}}-1)$ that also include the measurement noise, a controller $\kappa$ and a finite number of simulation steps $N_{\text{sim}}$.
A closed-loop finite-time performance indicator is a measurable function $\phi(w; N_{\text{sim}}, \kappa):\mathcal{W} = \mathbb{R}^{n_x} \times \mathbb{R}^{n_x} \times \mathbb{R}^{n_d}\times \dots \times \mathbb{R}^{n_d} \rightarrow \mathbb{R}$ that takes as input all variables defining the closed-loop trajectories for a controller $\kappa$ until time $N_{\text{sim}}$ and gives a scalar as a measure of closed-loop performance:
\begin{align}
    \phi(w; N_{\text{sim}}, \kappa) = \phi(x(0),\hat x(0), d(0), d(1), \dots, d(N_{\text{sim}}-1)).
\end{align}
\end{definition}
\begin{assumption}\label{assumption:simulation}
There exists a simulator that is able to compute closed-loop trajectories defined by~\eqref{eq:closed-loop}. In addition, there exists a known operator $\Phi_k$ that provides the state estimation $\hat{x}(k+1)$ from $\hat{x}(k)$, $y(k)$ and $u(k)$. That is, 
\begin{align}
    \hat x(k+1) = \Phi_k(\hat x(k), y(k),u(k)).
\end{align}
\end{assumption}
Assumption~\ref{assumption:simulation} implies that given $N_{\text{sim}}$ and the controller $\kappa$, the closed-loop trajectories are completely determined by $w$.
Probabilistic validation normally relies on a binary performance indicator that determines if the closed-loop is admissible or not. That is,
$$\phi(w; N_{\text{sim}},\kappa)=\bsis{rl} 0 &\mbox{ if the closed-loop trajectory is admissible for } w, \\ 1 &  \mbox{otherwise}. \esis$$
For this particular setting, one can resort to well-known results to obtain probabilistic guarantees about the probability $\Pr_\setW(\phi(w,N_{\text{sim}},\kappa))$. For a review on these results, see \cite{Alamo:15}. See also \cite{hertneck2018learning}, where  Hoeffding's inequality \cite{Hoeffding:63} is used to derive probabilistic guarantees in the context of learning an approximate model predictive controller. 

In this paper we address a more general setting in which we do not circumscribe the performance indicator to the class of binary functions. For example, we consider the closed-loop finite-time performance indicator given by the largest value of any component of $\constraint_l$ along the closed-loop simulation:
\begin{align}
    \phi(w; N_{\text{sim}}, \kappa) = \max_{\substack{k = 0, \dots, N_{\text{sim}}-1\\ l=1, \dots, n_\constraint}} \constraint_l(x(k), \kappa(\hat x(k)), d(k)).
\end{align}
Another possibility is to consider the average constraint violation as a performance indicator. That is,
\begin{align}\label{equ:average:violation}
    \phi(w; N_{\text{sim}}, \kappa) = \frac{1}{N_{\text{sim}}  n_{\text{g}}}\sum_{k = 0}^{N_{\text{sim-1}}} \sum_{l = 1}^{n_\constraint} \max\{0, \constraint_l(x(k), \kappa(\hat x(k)), d(k))\}.
\end{align}
Moreover, in many applications it is relevant to consider  indicators related to the closed-loop cost, such as an average cost:
\begin{align}
    \phi(w; N_{\text{sim}}, \kappa) = \frac{1}{N_{\text{sim}}}\sum_{k = 0}^{N_{\text{sim-1}}} \ell(x(k), \kappa(\hat x(k))),
\end{align}
or any other combination. In the following section we address how to obtain probabilistic guarantees on the random variable $\phi(w; N_{\text{sim}}, \kappa)$.


\section{Probabilistic validation} \label{sec:validation}
 
The derived closed-loop performance indicators can be used in the framework of probabilistic validation \cite{Tempo13,Alamo:15} to obtain probabilistic guarantees regarding the satisfaction of a given set of control specifications. In this section we present a novel result that allows us to address the probabilistic validation of arbitrary control schemes where the performance is influenced by hyper-parameters, e.g. backoff parameters or the control sampling time. 

\subsection{Probabilistic performance indicator levels}

We consider a finite family of controllers
$$ \kappa_i(\hat x),\; i=1,\ldots,M, $$ corresponding to $M$ different combinations of hyper-parameter values, e.g. for constraint backoffs or control sampling times. The objective of this section is to provide a probabilistic validation scheme that allows us to choose, from the $M$ possible controllers, the one with the  best  probabilistic certification for any given closed-loop finite-time performance indicator $\phi(w ; N_{\text{sim}}, \kappa_i)$. For simplicity in the notation, we denote the closed-loop finite-time performance indicator obtained with the controller $\kappa_i$ with $N_{\text{sim}}$ simulation steps as $\phi_i(w)$.


\begin{remark}\label{assumption:sampling}
The stochastic variable $w$ that defines the closed-loop trajectories follows the probability distribution $\setW$ from which it is possible to obtain independent identically distributed (i.i.d.) samples.
\end{remark}
Remark~\ref{assumption:sampling} only requires knowledge of the probability distributions of the uncertainty.

\begin{definition}[Probabilistic performance indicator level]
We say that $\gamma\in \R$ is a probabilistic performance indicator level with violation probability $\epsilon \in (0,1)$ for a sample $w$ drawn from $\setW$ for the measurable function $\phi:\setW \to \R$ if the probability of violation $\Pr_{\setW}\{ \cdot \}$ satisfies
$$ \Pr_{\setW} \{ \phi(w) > \gamma \} \leq \epsilon.$$
\end{definition}
To obtain probabilistic performance indicator levels for the considered controllers $\kappa_i$, $i=1,\ldots,M$, we generate $N$ i.i.d. scenarios 
$$w^{(j)} = \{x^{(j)}(0), \hat{x}^{(j)}(0), d^{(j)}(0), \dots, d^{(j)}(N_{\text{sim}-1})\}, \; j=1,\ldots,N.$$ 

For a given controller $\kappa_i$, with $1\leq i\leq M$, and the uncertain realizations $w^{(j)}$, $j=1,\ldots,N$, one could simulate the closed-loop dynamics and obtain the performance indicator corresponding to each uncertain realization. That is, one could obtain
$$\vv{v}_i= [\phi_i(w^{(1)}),  \phi_i(w^{(2)}), \ldots, \phi_i(w^{(N)})]\T \in \R^{N}.$$ 
It is clear that the largest value of the components of $\vv{v}_i$ could serve as an empirical performance level for the controller $\kappa_i$ provided that $N$ is large enough \cite{TeBaDa:97}. Another possibility is to discard the $r-1$ largest components of $\vv{v}_i$ and consider the largest of the remaining components as a (less conservative) empirical performance indicator level ($r$ equal to one corresponds to not discarding any component) \cite{Alamo:18}.  In the following section we show how to choose $N$ such that the obtained empirical performance indicator levels are, with high confidence $1-\delta$, probabilistic performance indicator levels with probability of violation $\epsilon$.

\subsection{Sample complexity}\label{section:generalization:max:function}

We first present a generalization of the notion of the maximum of a collection of scalars. This generalization is borrowed from the field of order statistics \cite{Ahsanullah:13}, \cite{Arnold:92}, and will allow us to reduce the conservativeness that follows from the use of the standard notion of max function. See also Section 3 of \cite{Alamo:18}.

\begin{definition}[Generalized max function]
Given the vector $$\vv{v} = [ v^{(1)},v^{(2)}, \ldots, v^{(N)}]\T\in \R^N,$$  and the integer $r$ with $1\leq r\leq N$ we denote
$$\psi(\vv{v},r)=v_+^{(r)},$$
where the vector $\vv{v_+} = [ v_+^{(1)},v_+^{(2)}, \ldots, v_+^{(N)}]\T\in \R^N$ is obtained by rearranging the values of the components of $\vv{v}$ in a non-increasing order. That is,
$$ v_+^{(1)}\geq v_+^{(2)} \geq \ldots \geq v_+^{(N-1)} \geq v_+^{(N)}.$$
\end{definition}

Clearly, given $\vv{v} = [ v^{(1)},\ldots, v^{(N)}]\T\in \R^N,$ we have $$ \psi(\vv{v},1) = \max\limits_{1\leq i\leq N}\,v^{(i)}, \;\; \psi(\vv{v},N) = \min\limits_{1\leq i\leq N}\,v^{(i)}.$$
Furthermore,  $\psi(\vv{v},2)$ denotes the second largest value in $\vv{v}$, $\psi(\vv{v},3)$ the third largest one, etc. We notice that the notation $\psi(\vv{v},r)$ does not need to make explicit $N$ and the number of components of $\vv{v}$. 

The next theorem provides a way to compute probabilistic performance levels for a family of $M$ controllers. 
The theorem constitutes a generalization of a similar result, presented in \cite{Alamo:18} for the particular case $M=1$. See also the seminal paper \cite{TeBaDa:97} for the particularization of the result to the case $r=1$, $M=1$.

\begin{theorem}\label{theo:probabilistic}
Given the controllers $\kappa_i$, $i=1,\ldots,M$, and the integer  $r\geq 1$, suppose that $N$ i.i.d. scenarios $$w^{(j)} = \{x^{(j)}(0), \hat{x}^{(j)}(0), d^{(j)}(0), \dots, d^{(j)}(N_{\text{sim}}-1)\}, \; j=1,\ldots,N,$$
are generated. We denote with $\vv{v}_i$, $i=1,\ldots,M$, the vector of performance indicators corresponding to the controller $\kappa_i$. That is, 
$$\vv{v}_i= [\phi_i(w^{(1)}),  \phi_i(w^{(2)}), \ldots, \phi_i(w^{(N)})]\T \in \R^{N}, \; i=1,\ldots, M.$$
Then, with probability no smaller than $1-\delta$, we have a probability of violation
$$ \Pr_\setW \{ \phi_i(w) > \psi(\vv{v}_i,r)\} \leq \epsilon, \; i=1,\ldots,M,$$
provided that $1\leq r \leq N$ and 
\begin{align}\label{eq:binomial_ineq}
  \Sum{\zeta=0}{r-1} \conv{N}{\zeta}\epsilon^\zeta (1-\epsilon )^{N-\zeta} \leq \frac{\delta}{M}.   
 \end{align}
 In addition,  \eqref{eq:binomial_ineq} is satisfied if:
 \begin{align}\label{eq:number_samples}
N \geq \frac{1}{\epsilon}\left(  r-1 + \ln\frac{M}{\delta} + \sqrt{2(r-1)\ln \frac{M}{\delta}}\right).     
 \end{align}
 \end{theorem}

\proof 
Given the controller $\kappa_i$ and $\gamma\in \R$, we denote $E_i(\gamma)$ the probability of the event $\phi_i(w)>\gamma$. That is,
$$ E_i(\gamma) \coloneq \Pr_{\setW}\{ \phi_i(w)>\gamma\}.$$
We denote probability of asymptotic failure the probability of generating $N$ i.i.d scenarios and obtaining an empirical probabilistic performance level that does not meet the probabilistic specification on probability of violation. We now make use of Property 3 in \cite{Alamo:18}, which states that, with probability no smaller than 
$$ 1- \Sum{\zeta=0}{r-1} \conv{N}{\zeta}\epsilon^\zeta (1-\epsilon )^{N-\zeta},$$
we have $$ E_i(\psi(\vv{v}_i,r))= \Pr_{\setW}\{ \phi_i(w)>\psi(\vv{v}_i,r)\} \leq \epsilon.$$
This means that the probability of asymptotic failure $\Pr_{\setW^N}\{ \cdot \}$ for $N$ samples $w^{(i)},\dots,w^{(N)}$ drawn from $\setW$ satisfies
$$ \Pr_{\setW^N} \{ E_i(\psi(\vv{v}_i,r)) > \epsilon \} \leq \Sum{\zeta=0}{r-1} \conv{N}{\zeta}\epsilon^\zeta (1-\epsilon )^{N-\zeta} \coloneq B(N,\epsilon,r-1). $$
 Consider now the probability $\delta_F$ that, after drawing $N$ i.i.d. samples $w^{(j)}$, $j=1,\ldots,N$, one or more of the empirical performance indicator levels $$ \gamma_i=\psi(\vv{v}_i,r),\; i=1,\ldots, M,$$ are not probabilistic performance indicator levels with violation probability $\epsilon$.  
 We have 
 \begin{eqnarray*}
 \delta_F  &=&  \Pr_{\setW^N} \{ \max\limits_{i=1,\ldots,M} E_i(\psi(\vv{v}_i,r)) > \epsilon   \} \\ 
 & \leq & \Sum{i=1}{M}\Pr_{\setW^N} \{ E_i(\psi(\vv{v}_i,r)) > \epsilon \} \\ 
 & \leq & M \Sum{\zeta=0}{r-1} \conv{N}{\zeta}\epsilon^\zeta (1-\epsilon )^{N-\zeta} \leq \delta. 
 \end{eqnarray*}
 That is, $\delta_F$ is smaller or equal than $\delta$ provided that  (\ref{eq:binomial_ineq}) holds. 
 This proves the first claim of the property. The second one follows directly from Corollary 1 of \cite{Alamo:15}, which provides an explicit number $N$ of samples that guarantees that a binomial expression $B(N,\epsilon,r-1)$ is smaller than a given constant. 
 \QED

The major advantage of Theorem~\ref{theo:probabilistic} is that a family of controllers can be evaluated for the same $N$ samples.
This is beneficial when the family of controllers can be evaluated in parallel or when drawing samples is expensive, e.g. in experimental setups.
The number of required samples for the same probabilistic statement is significantly smaller as when all controllers would be evaluated in a sequential approach as in~\cite{oishi2007polynomial,calafiore2011research,alamo2015randomized}.

\begin{remark}
Given a family of controllers $\kappa_i$, $i=1,\ldots,M$, one does not need to compute all the empirical performance indicator levels $\psi(\vv{v}_i,r)$, $i=1,\ldots,M$. It is sufficient to find one that meets the desired performance indicator levels. For example, if the performance indicator $\phi_i(w)$ is defined as the average constraint violation along the trajectory (see (\ref{equ:average:violation})), then the controller $\kappa_i$ provides an admissible closed-loop trajectory for $w$ if and only if $\phi_i(w)=0$.
In this case, the empirical performance indicator $\psi(\vv{v}_i,r)$ corresponding to $N$ i.i.d. scenarios is equal to 0 if no more than $r-1$ trajectories are non-admissible when applying the controller $\kappa_i$ to the scenarios. If $N$ is chosen according to (\ref{eq:binomial_ineq}) then Theorem \ref{theo:probabilistic} implies that with probability no smaller than $1-\delta$, all the controllers $\kappa_i$, $i=1,\ldots,M$,  providing $\psi(\vv{v}_i,r)=0$ are such that 
$$ \Pr_{\setW} (\phi_i(w)>0) \leq \epsilon. $$
It is also important to remark that the cardinality $M$ of the family of proposed controllers has little effect on the sample complexity $N$ because it appears into a logarithm. See also Subsection 4.2 in \cite{Alamo:15} for other randomized approaches based on a design space of finite cardinality.  
\end{remark}

\begin{remark}
Theorem~\ref{theo:probabilistic} can also be applied in the case when the performance indicators only take binary values. This has been presented in a similar form in~\cite{Alamo10ACC} and was used for control design problems. See, for example,~\cite{alamir2018feedback},~\cite{Alamir:15}.
\end{remark}

\section{Robust output-feedback nonlinear model predictive control}\label{sec:robust}

In this work our goal is to design an NMPC scheme that is able to control the uncertain nonlinear system~\eqref{eq:LTI} in an output-feedback setting where not all the states can be measured as described by the output equation~\eqref{eq:meas_function}. While the novel probabilistic validation scheme described in Section~\ref{sec:validation} can be applied using any controller, we believe that because of the complexity of the general robust output-feedback problem, it is a promising idea to use an approximate robust NMPC scheme that is validated a posteriori using probabilistic validation.

There exist many different robust model predictive control schemes, but there are four important characteristics that differentiate one approach from the other: the choice of cost function, the propagation of the uncertainty, robust constraint satisfaction and the characterization of feedback information.

The \emph{cost function} can be chosen following a min-max approach, where the worst-case realization of the uncertainty $d(k)$ at each step in the prediction is chosen~\cite{campo1987}. Tube-based methods usually choose the cost incurred by the closed-loop system driven by the nominal realization of the uncertainty~\cite{rawlings2009}. Scenario-tree based methods use a weighted sum of a set of discrete scenarios~\cite{lucia2013} and stochastic MPC schemes~\cite{delaPena2005b} make  use of, e.g., the expectation operator. In this work, we consider a general cost function $V (\mathcal{X}(0),\mathcal{D};N_{\text{p}},\kappa)$ that depends on an initial state set $\mathcal{X}(0)$, the uncertainty set $\mathcal{D}$, the prediction horizon $N_{\text{p}}$ and the control policy $\kappa$.

The \emph{propagation of the uncertainty} is one of the key elements of any robust NMPC scheme. A general framework, which is used in this work, is the definition of reachable sets at each sampling time in the prediction based on a current initial condition, the system model, the applied input and the uncertainty set $\mathcal D$. The reachable set at sampling time $k+1$  can be thus denoted as:
\begin{align}\label{eq:reachable}
\mathcal{X}(k+1) = \{& f(x(k), u(k),d(k)): x(k)\in \mathcal{X}(k),  d(k) \in \mathcal{D}\}.
\end{align}
The are several methods to compute such reachable sets. In the linear case, the consideration of the vertices of the uncertain set and their propagation along the prediction horizon is enough to compute an exact reachable set. In the nonlinear case, linearization techniques~\cite{althoff2014} or ODE bounding techniques~\cite{sahlodin11} can be used to obtain guaranteed over-approximations, which can be then used in robust optimal control schemes. 
To maintain the notation independent of the method used to obtain an (over-) approximation of the reachable sets at each sampling time, the \emph{bounding} operator denoted as $\diamond f(\cdot)$ is used, which is defined as:
\begin{align}\label{eq:reachable}
\mathcal{X}(k+1) &= \diamond f(\mathcal{X}(k), u(k), \mathcal{D} ).
\end{align}
Another possibility for the propagation of uncertainty is to resort to probabilistic reachable sets as done in \cite{hewing2018cdc}.

\emph{Robust constraint satisfaction} is often one of the main motivations for the use of a robust NMPC approach. It means that the requirements of the closed-loop system in form of input and state constraints should be satisfied for all possible outcomes of the uncertainty and it is usually enforced by embedding the reachable sets~\eqref{eq:reachable} into the constraints of an optimization problem.

The \emph{characterization of feedback} that is employed is another key property of any robust MPC scheme. It is well known that considering a sequence of optimal control inputs in the prediction under uncertainty can result in very conservative performance of the closed-loop because it is ignored that new information about the future will be available in the form of measurements and thus future actions can be adapted accordingly. To avoid this conservatism, closed-loop approaches can be used, in which one optimizes over a sequence of control policies $\kappa$ and can be formulated as the receding horizon solution of the following optimization problem:

\begin{subequations}\label{prob:MPC_ideal}
\begin{align}
& \underset{\kappa(\cdot)}{\text{minimize}}
& & \hspace{-2cm} V (\mathcal{X}(0),\mathcal{D};N_{\text{p}},\kappa), \label{eq:cost_mpc}\\
\mathbb{P}_{\text{ideal}}:\hspace{1.5cm} & \text{subject to}
& & \hspace{-2cm} {\mathcal{X}}(k+1) = \diamond f({\mathcal{X}}(k), \kappa (\mathcal X(k)), \mathcal D ), \quad \text{for }  k = 0, \ldots, N_{\text{p}}-1,\\
&&& \hspace{-2cm} \constraint_{l}({\mathcal{X}} (k), \kappa({\mathcal{X}} (k)) , \mathcal D) \leq 0, \;l=1,\ldots,n_\constraint, \quad \text{for }  k = 0, \ldots, N_{\text{p}}-1, \label{eq:cons_ideal}\\
&&& \hspace{-2cm} \mathcal X (N_{\text{p}}) \subseteq \mathcal{X}_f ,\\
&&& \hspace{-2cm} \mathcal X (0) = \{\hat x(0)\} \oplus \mathcal E_{\text{est}},
\end{align}
\end{subequations}
where the constraints~\eqref{eq:cons_ideal} denote that $\constraint_{l}(x, \kappa(x), d) \leq 0$ should be satisfied for all $x\in \mathcal X(k)$ and for all $d \in \mathcal D$.


Solving the ideal robust NMPC problem $\mathbb{P}_{\text{ideal}}$ defined in~\eqref{prob:MPC_ideal}, one obtains a receding horizon policy $\kappa_{\text{ideal}}(\hat x(0))$ which is a function of the initial state estimate $\hat x(0)$ that has been obtained with a certain estimation error bounded by $\mathcal E_{\text{est}}$.

Obtaining an exact solution of $\mathbb{P}_{\text{ideal}}$ is usually intractable mainly because of the bounding operator $\diamond f(\cdot)$ and the general feedback law $\kappa(\cdot)$. There are different alternative solutions to obtain approximations of this problem. A common simplifying assumption is to restrict the search to affine policies on the state or on the disturbances~\cite{goulart2006}. A different alternative is the use of a scenario tree~\cite{scokaert1998}, \cite{delaPena2005b}, \cite{lucia2013} in a so-called multi-stage NMPC approach. A multi-stage NMPC scheme is based on the representation of the uncertainty via a scenario tree (see Figure~\ref{fig:scenario_tree}), which branches at each sampling time.
This means that the uncertainty set is approximated by a discrete number of uncertainty realizations:
\begin{align}
    \mathcal{D} \approx \tilde{\mathcal{D}}  = \{d_1,\dots, d_s\}
\end{align}
where $s$ is the number of possible realizations of the uncertainty that are considered in the tree.
The considered realizations mean that each node branches $s$ times which results in $s^k$ nodes at stage $k$.
Using a scenario tree formulation, an approximation of the reachable set can be obtained as the convex hull of the set of all the nodes at a given stage, i.e.:
\begin{align}
    \mathcal X(k) \approx \text{Conv}(\tilde{\mathcal X}(k)) = \text{Conv}\left(\bigcup\limits_{i=1}^{s^k}  x_i(k)\right),
\end{align}
where $\text{Conv}(\cdot)$ denotes the convex hull of a set and $x_i(k)$ denotes the node $i$ of the tree at stage $k$ as depicted in Figure~\ref{fig:scenario_tree}. In the linear case with polytopic uncertainty, including the extreme values of the uncertainty in $\tilde{\mathcal{D}}$  guarantees an exact representation of the actual reachable set. In the nonlinear case considered in this paper it is only an approximation and therefore we focus on the point-wise approximation $\tilde{\mathcal{X}}$. Following the same notation, the bounding operator used to propagate the point-wise uncertainty description can be denoted as:

\begin{align}\label{eq:reachabe_ms}
\tilde{\mathcal{X}}(k+1) = \diamond f(\tilde{\mathcal{X}}(k), \kappa (\tilde{\mathcal X}(k)), \tilde{\mathcal D} ) \approx \bigcup\limits_{i=1}^{s^k} \bigcup\limits_{j=1}^{s} f(x_i(k), u_i(k), d_j).
\end{align}
The cost function is usually chosen as a weighted sum of the stage cost for each node in the scenario tree:
\begin{align}\label{eq:cost_ms}
V (x_1(0),\tilde{\mathcal{D}};N_{\text{p}},\kappa) = \sum_{k=0}^{N_{\text{p}}-1}\sum_{i = 1}^{s^k}\ell(x_i(k), \kappa(x_i(k))) + \sum_{i=1}^{s^{N_{\text{p}}}}V_f(x_i(N_{\text{p}})).
\end{align}

\begin{figure}[t]
\begin{center}
\tikzstyle{level 1}=[level distance=3.5cm, sibling distance=3.5cm]
\tikzstyle{level 2}=[level distance=3.5cm, sibling distance=2cm]

\tikzstyle{bag} = [circle, minimum width=4pt,fill, inner sep=0pt]
\tikzstyle{end} = [circle, minimum width=4pt,fill, inner sep=0pt]

\begin{tikzpicture}[grow=right,line width=0.3mm]
\node[bag, label = below: {$x_1(0)$}]{}
    child {
        node[bag, label = below: {$x_2(1)$}]{}
            child {
                node[end, label=below:
                    {$x_4(2)$}] {}
                edge from parent
                node[above] {}
                node[below=0.1cm]  {$u_2(1), d_2$}
            }
            child {
                node[end, label=below:
                    {$x_3(2)$}] {}
                edge from parent
                node[above=0.1cm] {$u_2(1), d_1$}
                node[below]  {}
            }
            edge from parent 
            node[above] {}
            node[below=0.2cm]  {$u_1(0), d_2$}
    }
    child {
        node[bag, label = below: {$x_1(1)$}] {}    
        child {
                node[end, label=below:
                    {$x_2(2)$}] {}
                edge from parent
                node[above] {}
                node[below=0.1cm]  {$u_1(1), d_2$}
            }
            child {
                node[end, label=below:
                    {$x_1(2)$}] {}
                edge from parent
                node[above=0.1cm] {$u_1(1), d_1$}
                node[below]  {}
            }
        edge from parent         
            node[above=0.2cm] {$u_1(0), d_1$}
            node[below]  {}
    };
    
\node[below] at (0, -3.8){$\tilde{\mathcal X}(0) = \{x(0)\}$;};
\node[below] at (3.2, -3.8){$\tilde{\mathcal X}(1) = \{x_1(1), x_2(1)\}$;};
\node[below] at (7.3, -3.8){$\tilde{\mathcal X}(2) = \{x_1(2),\dots,  x_4(2)\}$;};
\end{tikzpicture}    
\caption{Scenario tree representation.}  
\label{fig:scenario_tree}                                 
\end{center}                                 
\end{figure}

Introducing~\eqref{eq:cost_ms} and~\eqref{eq:reachabe_ms} in the ideal formulation of robust NMPC $\mathbb{P}_{\text{ideal}}$ we obtain the optimization problem that should be solved at each sampling time: 
\begin{subequations}\label{prob:MPC_ms}
\begin{align}
& \underset{\kappa(\cdot)}{\text{minimize}} 
& & \hspace{-2cm}\sum_{k=0}^{N_{\text{p}}-1}\sum_{i = 1}^{s^k}\ell(x_i(k), \kappa(x_i(k))) + \sum_{i=1}^{s^{N_{\text{p}}}}V_f(x_i(N_{\text{p}})), \label{eq:cost_mpc}\\
\mathbb{P}_{\text{ms}}:\hspace{1.5cm} & \text{subject to}
& & \hspace{-2cm} \tilde{\mathcal{X}}(k+1) =  \bigcup\limits_{i=1}^{s^k} \bigcup\limits_{j=1}^{s} f(x_i(k), \kappa(x_i(k)), d_j), \quad \text{for }  k = 0, \ldots, N_{\text{p}}-1,\\
&&& \hspace{-2cm} \constraint_{l}(\tilde{\mathcal{X}} (k), \kappa(\tilde{\mathcal{X}} (k)), \tilde{\mathcal{D}}) \leq 0, \;l=1,\ldots,n_\constraint, \quad \text{for }  k = 0, \ldots, N_{\text{p}}-1,\label{eq:cons_ms}\\
&&& \hspace{-2cm} \tilde{\mathcal{X}} (N_{\text{p}}) \subseteq \mathcal{X}_f ,\\
&&& \hspace{-2cm} \tilde{\mathcal{X}} (0) = \{\hat x(0)\},
\end{align}
\end{subequations}
where the constraints~\eqref{eq:cons_ms} denote that $\constraint_{l}(x, \kappa(x), d) \leq 0$ should be satisfied for all $x\in \tilde{\mathcal X}(k)$ and for all $d \in \tilde{\mathcal D}$. 
The optimal solution of~\eqref{prob:MPC_ms} is denoted as the multi-stage NMPC feedback policy $\kappa_{\text{ms}}$.

To avoid the exponential growth of the tree with the prediction horizon, a usual additional simplifying assumption is to consider that the tree branches only up to a given stage (usually called robust horizon). While this simplification introduces further errors in the approximation of the reachable sets at each stage, it achieves good results in practice \cite{lucia2014e}.
The current estimation error as well as the presence of future estimation errors should be also included in the problem formulation to achieve stability and recursive feasibility guarantees. This can be done in a multi-stage framework as shown in~\cite{subramanian2018synergistic}, but additional uncertainties should be included in the scenario tree. To mitigate the exponential growth of the scenario tree with the number of considered uncertainties, we do not consider the estimation error directly in the formulation of the tree. 
Following ideas from tube-based MPC, these additional errors will be taken into account by means of constraint tightening as explained in Section~\ref{sec:deep_learning_approx}.

\section{Deep learning-based approximate robust NMPC}
\label{sec:deep_learning_approx}

Despite recent advances in algorithms and hardware, solving the simplified output-feedback robust NMPC problem defined in~\eqref{prob:MPC_ms} in real time can be challenging.
To avoid the need for the real-time solution of non-convex optimization problems, this work considers the data-based approximation of the implicit feedback law defined by~\eqref{prob:MPC_ms} following the same ideas as explicit model predictive control. 
Approximating an NMPC controller with a neural network was already proposed by~\cite{parisini1995} back in 1995, where the use of shallow networks (with only one hidden layer) was proposed. This suggestion is based on the universal approximation theory that shows that a neural network with only one layer can approximate any function with any desired accuracy under mild conditions~\cite{barron1993universal}. 

\subsection{Deep neural networks}

 The function approximators chosen for this work are Deep neural networks (DNNs). This is motivated by recent theoretical results that support the increased representation power of neural networks with several hidden layers as opposed to classical shallow networks~\cite{safran2017}.
For the approximation of MPC laws via deep neural networks good results were obtained in~\cite{Chen2018,zhang2019safe,hertneck2018learning,karg2018efficient} among other recent works.
In the case of linear time-invariant systems, it was shown in~\cite{karg2018efficient} that a deep neural network with a given size can exactly represent the explicit MPC law.
The robust NMPC problem~\eqref{prob:MPC_ms} is a parametric optimization problem that depends on the current (estimated) state and on the uncertainty values used to define the scenario tree. To perform a deep learning-based approximation, a finite amount of $N_{\text{s}}$ samples $x^{(i)}$ of the state space are chosen and then $N_{\text{s}}$ different optimization problems are solved to obtain the corresponding optimal inputs $\kappa_{\text{ms}}(x^{(i)})$.

A standard deep feed-forward neural network with fully connected layers is defined as a sequence of layers which determines a function $\mathcal{N}:\mathbb{R}^{n_x} \rightarrow \mathbb{R}^{n_u}$ of the form
\begin{align}\label{eq:neural_network}
\begin{split}
\mathcal{N}(x;\theta) = 
 \bigg \{
\begin{array}{lll}
		\alpha_{L+1} \circ \beta_L \circ \alpha_L \circ \dots \circ \beta_1 \circ \alpha_1(x) & \text{for} & L \geq 2, \\
		\alpha_{L+1} \circ \beta_1 \circ \alpha_1(x), & \text{for} & L = 1,
\end{array}
\end{split}
\end{align}
where the input of the network is $x \in \mathbb{R}^{n_x}$ and the output of the network is $ u \in \mathbb{R}^{n_u}$.
The dimensions of the network are defined by the number of hidden layers $L$ and the number of neurons $H$ per hidden layer, also denoted as the width of the hidden layer, when equal width for all hidden layers is assumed.
In contrast to \emph{shallow} neural networks with $L=1$ hidden layers, \emph{deep} neural networks have $L \geq 2$ hidden layers.
The complexity of a neural network can be defined either by the number of weights
\begin{align}\label{eq:number_weights}
N_{\text{weights}} = n_x \cdot (H+1) + (L-1) \cdot (H+1) \cdot H + H \cdot (n_u + 1),
\end{align}
or the number of neurons
\begin{align}\label{eq:number_neurons}
N_{\text{neurons}} = L \cdot H,
\end{align}
that form a given network.
The number of weights defines the necessary memory that is needed to store a neural network while the number of neurons determines the maximum possible amount of nonlinear functions present in the approximation.
Each hidden layer connects a preceding affine function:
\begin{align}\label{eq:affine_function}
\alpha_l(\xi_{l-1}) = W_l\xi_{l-1}+b_l,
\end{align}
where $\xi_{l-1} \in \mathbb{R}^H$ is the output of the previous layer and $\xi_0 = x$, with a nonlinear activation function $\beta_l$.
Common choices for the activation function are rectifier linear  units (\emph{ReLU}) and the sigmoid function or the hyperbolic tangent (\emph{tanh}):
\begin{align}
    \beta_l(\alpha_l) = \frac{e^{\alpha_l} - e^{-\alpha_l}}{e^{\alpha_l} + e^{-\alpha_l}},
\end{align}
which will be used throughout this work.
The parameters of all layers are summarized in $\lambda = \{\lambda_1, \dots, \lambda_{L+1}\}$ with
\begin{align}
\lambda_l = \{W_l,b_l\} \quad \forall l = 1, \dots, L+1,
\end{align}
where $W_l$ are the weights and $b_l$ are the biases describing the corresponding affine functions.
The best data-based approximation of the exact multi-stage NMPC~\eqref{prob:MPC_ms} with a neural network for a given training data set $\mathcal T = \{(x^{(1)},\kappa_{\text{ms}}(x^{(1)})),\dots, (x^{(N_{\text{s}})}, \kappa_{\text{ms}}(x^{(N_{\text{s}})}))\}$ with $N_{\text{s}}$ elements and fixed dimensions $L$ and $H$ is achieved for:
\begin{align}\label{eq:train}
\lambda^* = \underset{\lambda}{\argmin}\,\,\,\,\frac{1}{N_s}\sum_{i=1}^{N_{\text{s}}}(\kappa_{\text{ms}}(x^{(i)}) - \mathcal{N}(x^{(i)}; \lambda))^2.
\end{align}
The resulting deep learning-based controller is denoted as $\kappa_{\text{dnn}}(x) = \mathcal{N}(x;\lambda^*)$.

\subsection{Constraint tightening}

We propose to use a robust NMPC scheme to take explicitly into account the most important uncertainties that affect the system. Still, it is virtually impossible to account for all possible uncertainties and to obtain exact state estimates which in comparison to the ideal, robust NMPC feedback law $\kappa_{\text{ideal}}$ results in two sources of error:
\begin{align}\label{eq:errors_nmpc}
    ||\kappa_{\text{ideal}} (x(k)) - \kappa_{\text{ms}}(\hat x(k))|| \leq \epsilon_{\text{est}} + \epsilon_{\text{ms}},
\end{align}
where $\epsilon_{\text{est}}$ is the estimation and measurement error and $\epsilon_{\text{ms}}$ is the error caused by the approximation of the reachable set by a set of discrete scenarios.
Because solving the multi-stage NMPC problem~\eqref{prob:MPC_ms} online is challenging, our goal is to determine a candidate neural network controller by generating input-output data pairs via the solution of the multi-stage NMPC problem~\eqref{prob:MPC_ms} and approximating its solution via a deep neural network solving~\eqref{eq:train}. 
This means that the closed-loop will be controlled using the feedback law $\kappa_{\text{dnn}}$ that approximates the behavior of $\kappa_{\text{ms}}$ which introduces an error $\epsilon_{\text{approx}}$ on top of those described in~\eqref{eq:errors_nmpc}:
\begin{align}\label{eq:errors_dnn}
\begin{split}
    ||\kappa_{\text{ideal}} (x(k)) - \kappa_{\text{dnn}}(\hat x(k))|| &= ||\kappa_{\text{ideal}} (x(k)) - \kappa_{\text{ms}}(\hat x(k)) + \kappa_{\text{ms}}(\hat x(k)) - \kappa_{\text{dnn}}(\hat x(k))|| \\
    & \leq ||\kappa_{\text{ideal}} (x(k)) - \kappa_{\text{ms}}(\hat x(k))|| + ||\kappa_{\text{ms}}(\hat x(k)) - \kappa_{\text{dnn}}(\hat x(k))|| \\
    &\leq \epsilon_{\text{est}} + \epsilon_{\text{ms}} + \epsilon_{\text{approx}}.
\end{split}
\end{align}
Finding upper-bounds for each one of the errors to apply traditional robust NMPC schemes is not possible for the general nonlinear case. 

To counteract the possible errors $\epsilon_{\text{est}}$, $\epsilon_{\text{ms}}$, and $\epsilon_{\text{approx}}$, and following ideas from tube-based MPC, an additional backoff $\eta$ is used to tighten the original constraints of the robust NMPC problem that is solved to generate input-output data for training:

\begin{subequations}\label{prob:MPC_backoff}
\begin{align}
& \underset{\kappa(\cdot)}{\text{minimize}}
& & \hspace{-2cm}\sum_{k=0}^{N_{\text{p}}-1}\sum_{i = 1}^{s^k}\ell(x_i(k), \kappa(x_i(k))) + \sum_{i=1}^{s^{N_{\text{p}}}}V_f(x_i(N_{\text{p}})), \label{eq:cost_mpc}\\ 
\mathbb{P}_{\text{ms},\eta}:\hspace{1.5cm} & \text{subject to}
& & \hspace{-2cm} \tilde{\mathcal{X}}(k+1) =  \bigcup\limits_{i=1}^{s^k} \bigcup\limits_{j=1}^{s} f(x_i(k), \kappa(x_i(k)), d_j), \quad \text{for }  k = 0, \ldots, N_{\text{p}}-1,\\
&&& \hspace{-2cm} \constraint_{l}(\tilde{\mathcal{X}} (k), \kappa(\tilde{\mathcal{X}} (k)) , \tilde{\mathcal{D}}) \leq -\eta, \;l=1,\ldots,n_\constraint, \quad \text{for }  k = 0, \ldots, N_{\text{p}}-1,\\
&&& \hspace{-2cm} \tilde{\mathcal{X}} (0) = \{\hat x(0)\}.
\end{align}
\end{subequations}




Solving~\eqref{prob:MPC_backoff} online would lead to the feedback controller $\kappa_{\text{ms}}(\hat x, \eta)$. We are however interested in the proposed approximate robust NMPC $\kappa_{\text{dnn}}(\hat x, \eta)$ that is obtained by training a deep neural network via~\eqref{eq:train} based on input-output data generated by solving~\eqref{prob:MPC_backoff} for many different initial conditions.
Introducing a backoff $\eta$ does not guarantee in general that the closed-loop satisfies the constraints. For this reason, closed-loop constraint satisfaction is also not ensured a priori with a terminal set. 
The probabilistic design scheme presented in the previous sections is employed to select the backoff parameter $\eta$. The proposed methodology provides probabilistic guarantees on the performance indicators of the closed-loop uncertain system. 

\section{Case study}
\label{sec:kite_model}

We investigate the optimal control of a kite which is used to tow a boat.
The stable and safe operation of the kite is challenging due to the highly nonlinear system dynamics, uncertain parameters, strong influence from disturbances like wind speed and noisy measurements.
To develop optimal control schemes of a kite system, typically models with moderate complexity such as~\cite{houska2006,fagiano2010} are considered because of the required short sampling times.
Although for our proposed strategy, also a high-fidelity models could be considered since the majority of the computational load is shifted offline, we consider a popular three-state model as presented in \cite{erhard2013} to facilitate the comparison of the results with previous works.
We derive an approximate deep learning-based controller from a robust NMPC formulation, which enables a very fast and easy evaluation of the controller even on computationally limited hardware.
The idea of learning a controller for a kite has already been exploited in~\cite{fagiano2014automatic}, where polynomial basis functions were used to approximate the behaviour of a human pilot based on measurements.

\subsection{Kite model}
In the context of NMPC, we focus on the model presented in~\cite{erhard2013} which consists of three states, one control input and two uncertain parameters.
The state evolution is given by the ordinary differential equations of the three angles $\thetak$, $\phik$ and $\psik$ of the spherical coordinate system describing the position of the kite:

\begin{subequations}\label{eq_kite_model}
    \begin{align}
        \dot{\theta}_{\text{kite}} &=\frac{v_{\text{a}}}{L_{\text{T}}} (\cos \psik - \frac{\tan \thetak}{E}),\\
        \dot{\phi}_{\text{kite}} &= -\frac{v_{\text{a}}}{L_{\text{T}}\sin\thetak}\sin{\psik},\\
        \dot{\psi}_{\text{kite}} &= \frac{v_{\text{a}}}{L_{\text{T}}}\tilde{u} + \dot{\phi}_{\text{kite}} \cos{\thetak},
    \end{align}
where
\begin{align}
    v_{\text{a}} &= v_0 E \cos\thetak,\\
    E &= E_0 - \tilde{c}\tilde{u}^2.\label{eq_glide_ratio}
\end{align}
\end{subequations}
The angle between wind and tether (zenith angle) is described by \thetak, the angle between the vertical and the plane is denoted by  \phik and \psik represents the orientation of the kite.
The three states can be manipulated via the steering deflection $\tilde{u}$.
The area of the kite is denoted as $A$, and $L_{\text{T}}$ is the length of the tether.
The effect of the wind is denoted as $v_{\text{a}}$, which is strongly influenced by the wind speed $v_0$, the first uncertain parameter.
The glide ratio $E$ is dependent on the base glide ratio $E_0$, the second uncertain parameter, and the magnitude of the steering deflection $\tilde{u}$~\cite{costello2013}.
The parameters of the kite model are shown in the upper part of Table~\ref{tab:parameters}.

\subsection{Wind model}
The wind speed $v_{\text{0}}$ is considered as a single uncertainty in~\eqref{prob:MPC_backoff}, but the realizations of the values are computed based on a simulation model. 
The underlying wind model was presented in~\cite{costello2017crosswind} and is described by:
\begin{subequations}
    \begin{align}\label{eq:wind_model}
        v_0 = v_{\text{m}} + \bar{v}_{\text{N}} + \sigma_v c_v p_v,
    \end{align}
    where
    \begin{align}
        \sigma_v &= k_{\sigma_v} v_{\text{m}},\\
        \bar{v}_{\text{N}} &= -\sigma_v/(2v_{\text{m}}),\\
        \tau_{\text{F}} &= L_v/v_{\text{m}},\\
        K_{\text{F}} &= \sqrt{1.49\tau_{\text{F}}/T_v},\\
        c_v &= K_F/\tau_{\text{F}},\\
        \dot{p}_v &= -p_v/\tau_{\text{F}} + w_{\text{tb}},
    \end{align}
\end{subequations}
when the wind shear is neglected.
The term $w_{\text{m}}$ gives the current average wind speed, $w_{\text{tb}}$ is introduced as a white noise generator to model the short term turbulence and $p_v(0) = \normal(0,0.25)$ is the initial state of the turbulence, where $x_{\text{normal}} = \normal(\mu_{\text{normal}},\sigma_{\text{normal}})$ denotes that  the variable $x_{\text{normal}}$ follows a normal distribution with mean $\mu_{\text{normal}}$ and standard deviation $\sigma_{\text{normal}}$.
In a similar manner, $x_{\text{unif}} = \unif(a_{\text{unif}},b_{\text{unif}})$ means that the variable $x_{\text{unif}}$ follows a uniform distribution between $a_{\text{unif}}$ and $b_{\text{unif}}$.
An overview of the parameters for the wind model is given in the lower part of Table~\ref{tab:parameters}.
For further details on modeling assumptions and the choice of parameters the reader is referred to~\cite{costello2017crosswind}.

\subsection{Extended Kalman Filter}\label{subsec:extended_kalman_filter}
We assume that we can measure the two angles $\thetak$ and $\phik$ and the wind speed $v_0$.
An Extended Kalman Filter (EKF) is used to obtain an estimate of the augmented state $x_{\text{aug}} = [\thetak, \phik, \psik, E_0, v_0]^T$ in each control instance from the measurements:
\begin{align}\label{eq:measurement}
    y(x_{\text{aug}}) = [\thetak + w_{\thetak}, \phik + w_{\phik}, v_0 + w_{v_0}]^T,
\end{align}
with the zero-mean gaussian noises $w_{\thetak} = \normal(0,0.01)$, $w_{\phik} = \normal(0,0.01)$ and $w_{v_0} = \normal(0,0.05)$.
The augmented state is initialized for all simulations as $x_{\text{aug}}(0) = [\thetak(0) \cdot \delta_{\thetak}, \phik(0) \cdot \delta_{\phik}, \psik(0) \cdot \delta_{\psik}, E_0 \cdot \delta_{E_0}, v_0(0) \cdot \delta_{v_0}]^T$, where all noises $\delta_{(\cdot)}$  are drawn from $\normal(1,0.05)$. 
Neither the estimates of the uncertain parameters nor the measurement of the wind speed are used in the computations of the controller, because their possible values are considered in the robust NMPC approach.
The initial covariance matrix is given by $P_{\text{EKF}} = \text{diag}([1 \times 10^{-2}, 1 \times 10^{-2}, 1 \times 10^{-2}, 1.0, 2 \times 10^{-1})$, the estimate of the process noise by $Q_{\text{EKF}} = \text{diag}([1 \times 10^{-5}, 1 \times 10^{-5}, 1 \times 10^{-4}, 1 \times 10^{-5}, 3 \times 10^{-3}])$, the measurement noise matrix by $R_{\text{EKF}} = \text{diag}([1 \times 10^{-2}, 1 \times 10^{-2}, 5 \times 10^{-2}])$ and the observer has a sampling time of $t_{\text{EKF}} = \SI{0.05}{\second}$.

\subsection{Objective, constraints and control settings}
The goal of the control is to maximize the thrust of the tether defined by:
\begin{align}
T_{\text{F}} &= \frac{1}{2} \rho v_0^2 A\cos^2\thetak (E+1)\sqrt{E^2+1} \cdot(\cos\thetak \cos\beta + \sin \thetak\sin\beta\sin\phik),
\end{align}
while maintaining a smooth control performance and satisfying the constraints.
The desired behaviour is enforced in the stage cost:
\begin{align}\label{eq:stage_cost}
    \ell(x,u) = - w_{\text{F}} T_{\text{F}} + w_u (\tilde u - \tilde u_{\text{prev}})^2,
\end{align}
where $w_{\text{F}} = 1e-4$ and $w_u = 0.5$ are weights and $\tilde{u}_{\text{prev}}$ is the previous control input and sampling time of the controller $t_{\text{c}} = \SI{0.15}{\second}$ with a prediction horizon of $N_{\text{P}} = 40$ steps.

Throughout the operation of the kite it has to be ensured that the height of the kite:
\begin{align}\label{eq:height_kite}
h(x) = L_{\text{T}} \sin{\thetak} \cos{\phik},
\end{align}
never falls below $h_{\text{min}} = \SI{100}{\metre}$.
The height constraint is a critical constraint of the control task since the best performance is obtained when the kite is operated close to $h_{\text{min}}$.
Because of the error $\epsilon_{\text{ms}}$ caused by the approximation of the reachable sets in the multi-stage NMPC formulation, the errors due to a deep learning-based approximation $\epsilon_{\text{approx}}$ as well as the errors related to estimation and measurement errors  $\epsilon_{\text{est}}$, constraint satisfaction can not be guaranteed.
To cover the effect of the errors, the backoff parameter $\eta > \SI{0}{\metre}$ is introduced and the height constraint:
\begin{align}\label{eq:height_constraint}
    h(x) > h_{\text{min}} + \eta,
\end{align}
is formulated as a soft constraint to avoid numerical problems.

To build a multi-stage NMPC controller, we consider the combinations of the extreme values of the base glide ratio $E_0 \in [4,6]$ and the wind speed $v_0 \in [\SI{6}{\metre\per\second}, \SI{10}{\metre\per\second}]$ and a one-step robust horizon resulting in a total of four scenarios. The interval for the wind speed is obtained by summarizing the possible effects of the uncertain wind model parameters $v_{\text{m}}$, $p_v(0)$ and $w_{\text{tb}}$ into the single uncertain variable $v_0$.

\subsection{Simulation}
For the simulation of the system, it is assumed that the uncertain parameters $E_0$ and $w_{\text{m}}$ are constant over a given closed-loop simulation
and that $w_{\text{tb}}$ changes every $t_{\text{c}}=\SI{0.15}{\second}$.
The values of the uncertain parameters are drawn from the probability distribution described in Table~\ref{tab:parameters}.


\begin{table}
\caption{Overview of the model states and parameters and as which variable they are considered in~\eqref{prob:MPC_backoff}.}
\begin{center}
\begin{tabular}{c c c c  c c }\label{tab:parameters}
& \textbf{Symbol} &\textbf{Type} & \textbf{Values / Constraints} & \textbf{Units} & \textbf{Variable}\\
\cmidrule[0.3pt](){2-6}
 \parbox[t]{2.5mm}{\multirow{9}{*}{\rotatebox[origin=c]{90}{kite model}}} & $\theta_{\text{kite}}$         & State & $\interval{0}{\frac{\pi}{2}}$	          & \si{\radian} & $x$\\
& $\phi_{\text{kite}}$            & State & $\interval{-\frac{\pi}{2}}{\frac{\pi}{2}}$ & \si{\radian} & $x$\\
& $\psi_{\text{kite}}$            & State & $\interval{0}{2\pi}$	                      & \si{\radian} & $x$\\
& $\tilde{u}$     & Control input & $\interval{-10}{10}$	                      & \si{\newton} & $u$ \\
& $\tilde{c}$      & Known parameter & $0.028$	                                  & - & -\\
& $\beta$          & Known parameter & 0	                                      & \si{\radian} & - \\
& $\rho$           & Known parameter & 1	                                      & \si{\kilogram\per\cubic\metre} & - \\
& $h_{\text{min}}$           & Known parameter & 100	                                      & \si{\metre} & - \\
& $E_0$ 			 & Uncertain parameter & $\unif(4,6)$                          & - & $d$ \\
\cmidrule[0.1pt](lr){2-6}
 \parbox[t]{2.5mm}{\multirow{6}{*}{\rotatebox[origin=c]{90}{wind model}}}  & $p_v$  &  State     &  -                & \si{\second}  & \parbox[t]{2.5mm}{\multirow{6}{*}{\rotatebox[origin=c]{90}{$d$ via $v_0$}}} \\
& $k_{\sigma_v}$  &  Known parameter     &  0.14                & - & \\
& $L_v$           &  Known parameter     &  100                 & \si{\metre} & \\
& $T_v$           &  Known parameter     &  0.15                & \si{\second} & \\
& $v_{\text{m}}$  & Uncertain parameter  &  $\unif(7,9)$   &  \si{\metre\per\second} & \\
& $w_{\text{tb}}$ & Uncertain parameter  &  $\normal(0,0.25)$   &  - & \\
\cmidrule[0.3pt](){2-6}
\end{tabular}
\end{center}
\end{table}

\section{Results}\label{sec:results}

The proposed method for the probabilistic verification of controllers is analyzed for the towing kite case study.
The baseline controller for our investigations, which is also used for the training data generation for the corresponding approximate neural network controller $\kappa_{\text{dnn},\eta}$, is the exact multi-stage NMPC controller $\kappa_{\text{ms}}(\hat{x},\eta)$~\eqref{prob:MPC_backoff} that derives its initial state estimate $\hat{x}$ from the EKF based on the current measurement~\eqref{eq:measurement}.
This means that the baseline controller is affected by the estimation error $\epsilon_{\text{est}}$ and the error $\epsilon_{\text{ms}}$ caused by the discrete representation of the uncertainties in the scenario tree and hence no formal guarantees on constraint satisfaction can be given.
To avoid numerical problems for the solver in case of violations, the critical height constraint is implemented as a soft constraint.

\subsection{Learning an approximate output-feedback robust NMPC controller}


The training process of a neural network is determined by the quality of the data and the chosen hyperparameters like activation function of the hidden layers and network size~\eqref{eq:number_weights},~\eqref{eq:number_neurons}.
In the following, we discuss how the training data can be generated in a way that reduces the number of samples that are needed to achieve a satisfactory approximation in comparison to a random sampling.
For the training of the neural networks we used the toolbox Keras~\cite{chollet2015keras} with the backend TensorFlow~\cite{tensorflow2015-whitepaper} and Adam~\cite{kingma2014} as the optimization algorithm.
The weights were initialized based on the glorot uniform distribution~\cite{glorot2010understanding} and the biases were set to zero.
All considered networks use hyperbolic tangent (\emph{tanh}) as activation function in the hidden layers and a linear output layer.
As the focus of this work is the verification of a given approximate controller and not the training process or the choice of the optimal network architecture, we refrained from applying methods such as Bayesian Optimization to obtain an optimal structure of the underlying network~\cite{kandasamy2018neural}.


We consider two training data sets $\mathcal{T}_{\text{feas}}$ and $\mathcal{T}_{\text{opt}}$, and two validation data sets $\mathcal{V}_{\text{feas}}$ and $\mathcal{V}_{\text{opt}}$.
Each data data set contains samples $(x^{(i)},\kappa_{\text{ms}}(x^{(i)}))$ corresponding to the numerical solution of the multi-stage problem~\eqref{prob:MPC_backoff} at state $x^{(i)}$.
The subscript \emph{opt} indicates that the data was derived from optimal closed-loop trajectories, e.g. $\mathcal{T}_{\text{opt}} = \{(x^{(j)},\kappa_{\text{ms}}(x^{(j)})),\dots,(x^{(N_\text{sim} \cdot N_\text{traj})},\kappa_{\text{ms}}(x^{(N_\text{sim} \cdot N_\text{traj})}))\}$ is composed of $N_\text{traj}$ state-feedback closed-loop simulations of length $N_\text{sim}$ using the exact multi-stage NMPC~\eqref{prob:MPC_backoff} under the dynamics presented in~\eqref{eq_kite_model}, where the uncertain parameters of the model and the initial conditions are drawn according to the distributions given in Table~\ref{tab:parameters} and first row of Table~\ref{tab:uncertainty_sampling} respectively.
The subscript \emph{feas} means that the data was obtained at randomly sampled states, e.g. $\mathcal{T}_{\text{feas}} = \{(x^{(i)},\kappa_{\text{ms}}(x^{(i)})),\dots,(x^{(N_\text{s})},\kappa_{\text{ms}}(x^{(N_\text{s})})) \}$ is obtained by sampling $x^{(i)}$ uniformly from the feasible state space and solving~\eqref{prob:MPC_backoff}. 
Since the training data is generated based on simulations, the application of output-feedback via EKF is not necessary and not used for the data generation.
Each trajectory consists of $N_{\text{sim}} = 400$ simulation steps which results in a total simulation time of $t_{\text{sim}} = N_{\text{sim}} \cdot t_{\text{c}} = \SI{60}{\second}$.
For $\mathcal{T}_{\text{opt}}$, $N_{\text{traj}} = 200$ closed loop runs were simulated leading to $N_{\text{traj}} \cdot N_{\text{sim}} = 80000$ data pairs and for the validation $N_{\text{traj}} = 50$ simulations were rolled out, resulting in $N_{\text{traj}} \cdot N_{\text{sim}} = 20000$ samples in $\mathcal{V}_{\text{opt}}$.
For the data sets $\mathcal{T}_{\text{feas}}$ and $\mathcal{V}_{\text{feas}}$, $N_{\text{s}} = 80000$ and $N_{\text{s}} = 20000$ random samples were drawn respectively.

\begin{figure}
\begin{center}
\includegraphics[width=0.5\linewidth, trim=0.3cm 0.0cm 0.2cm 1.8cm, clip]{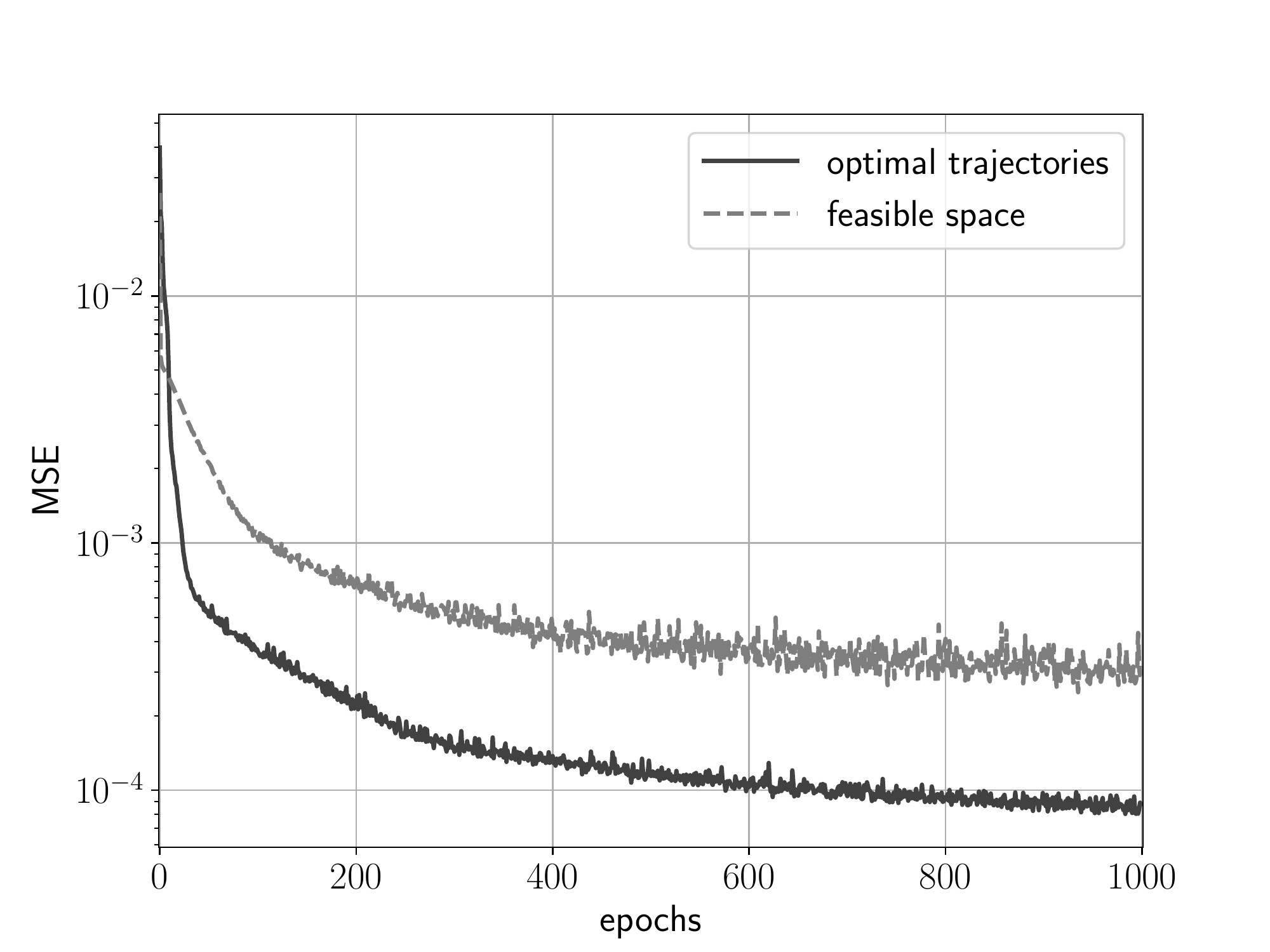}
\label{fig:opt_vs_feas}
\caption{Mean squared error obtained when training a deep neural network using the space of optimal trajectories $\mathcal{T}_{\text{opt}}$ or the full feasible space $\mathcal{T}_{\text{feas}}$ as training data.}\label{fig:training_process}
\end{center}
\end{figure}

For the following investigations, we trained five deep networks with $L = 6$ layers and $H = 30$ neurons per layer on each training set and evaluated all five obtained networks on each validation set.
By averaging the results over five networks the impact of the stochastic learning is reduced.
Training a deep neural network with the data pairs $\mathcal{T}_{\text{opt}}$ leads to a significantly smaller average mean squared error (MSE) when compared to the training using the training data $\mathcal{T}_{\text{feas}}$, as Figure~\ref{fig:training_process} shows, because the sampled space of optimal trajectories is smaller in comparison to the feasible space.
To investigate the impact of the training data set on the actual performance, the networks are tested on the validation sets $\mathcal{V}_{\text{feas}}$ and $\mathcal{V}_{\text{opt}}$.
The networks trained on $\mathcal{T}_{\text{feas}}$ perform better when evaluated on whole feasible space with an average MSE of 0.0048 in comparison to the networks trained on $\mathcal{T}_{\text{opt}}$ with an average MSE of 0.2105.
But when the networks are evaluated on the space of optimal closed-loop trajectories via $\mathcal{V}_{\text{opt}}$, 
the networks trained on $\mathcal{T}_{\text{opt}}$ have a significantly smaller average MSE of 0.0087 than networks trained on $\mathcal{T}_{\text{feas}}$ with an average MSE of 0.1642.
The fact that controllers trained on $\mathcal{T}_{\text{opt}}$ do not cover the whole feasible space is not critical since the learning-based controller will only operate in the neighborhood of optimal trajectories where a close approximation of the exact multi-stage NMPC is achieved.
Additionally, the controllers will be probabilistically validated, and this validation is completely independent of the data used for training.
Our experience shows that extracting training data from closed loop trajectories can significantly reduce the necessary number of training samples $N_{\text{s}}$ and the dimensions $L$ and $H$ of the neural network to obtain a desired approximation error of the deep learning-based controller in the critical domain.


For all the results presented in the remainder of the paper, we use deep neural networks with $L = 6$ and $H = 30$ which were trained on the space of optimal trajectories $\mathcal{T}_{\text{opt}}$ due to the observed superior approximation quality in the crucial regions of the state space.

\begin{table}
\caption{Overview of the the parameter sampling via uniform distribution, normal distribution, beta(2,5) and pareto(5) distribution and  results of evaluating the approximate controller $\kappa_{\text{dnn}}$ with $\eta=\SI{4}{\metre}$ for 1388 randomly drawn scenarios $w$. The measurement noise $w_{\text{meas}} =[w_{\thetak},w_{\phik},w_{\psik}]^T$, the initial state of the turbulence $p_v(0)=\normal(0,0.25)$, the white noise modelling the short term turbulences $w_{\text{tb}}=\normal(0,0.25)$ and the initialization of the estimation vector $x_{\text{aug}}(0)$ is for all scenario spaces identical.}
\begin{center}
\begin{tabular}{c c c c c c c c}\label{tab:uncertainty_sampling}
distribution & $\thetak(0)$ [\SI{}{\degree}] & $\phik(0)$ [\si{\degree}] & $\psik(0)$  [\si{\degree}] & $E_0$ [-] &  $v_{\text{m}}$ [\si{\metre\per\second}] & feasible traj. & $\psi(\mathbf{v},4)$ [\si{\metre}]\\
\midrule
    Uniform       & (28.0,30.0)  & (-10.0,10.0)  & (-2.0,2.0) & (4.0,6.0)  & (7.0,9.0)  & 1385/1388  & -0.316\\
    Normal        & (29.0,0.35)  & (0.0,3.5)     & (0.0,0.7)  & (5.0,0.35) & (8.0,0.35) & 1387/1388  & -0.739\\
    Beta          & (2.0,28.0)   & (20.0,-10.0)  & (4.0,-2.0) & (2.0,4.0)  & (2.0,7.0)  & 1387/1388  & -0.556\\
    Pareto        & (5.0,28.0) & (5.0,-10.0) & (5.0,2.0) & (5.0,4.5) & (5.0,7.5) & 1385/1388 & -0.037\\
\bottomrule
\end{tabular}
\end{center}
\end{table}

\begin{figure}
\begin{center}
\begin{subfigure}{0.49\textwidth}
\includegraphics[width=.99\linewidth, trim=0.0cm 0.0cm 0.0cm 0.0cm, clip]{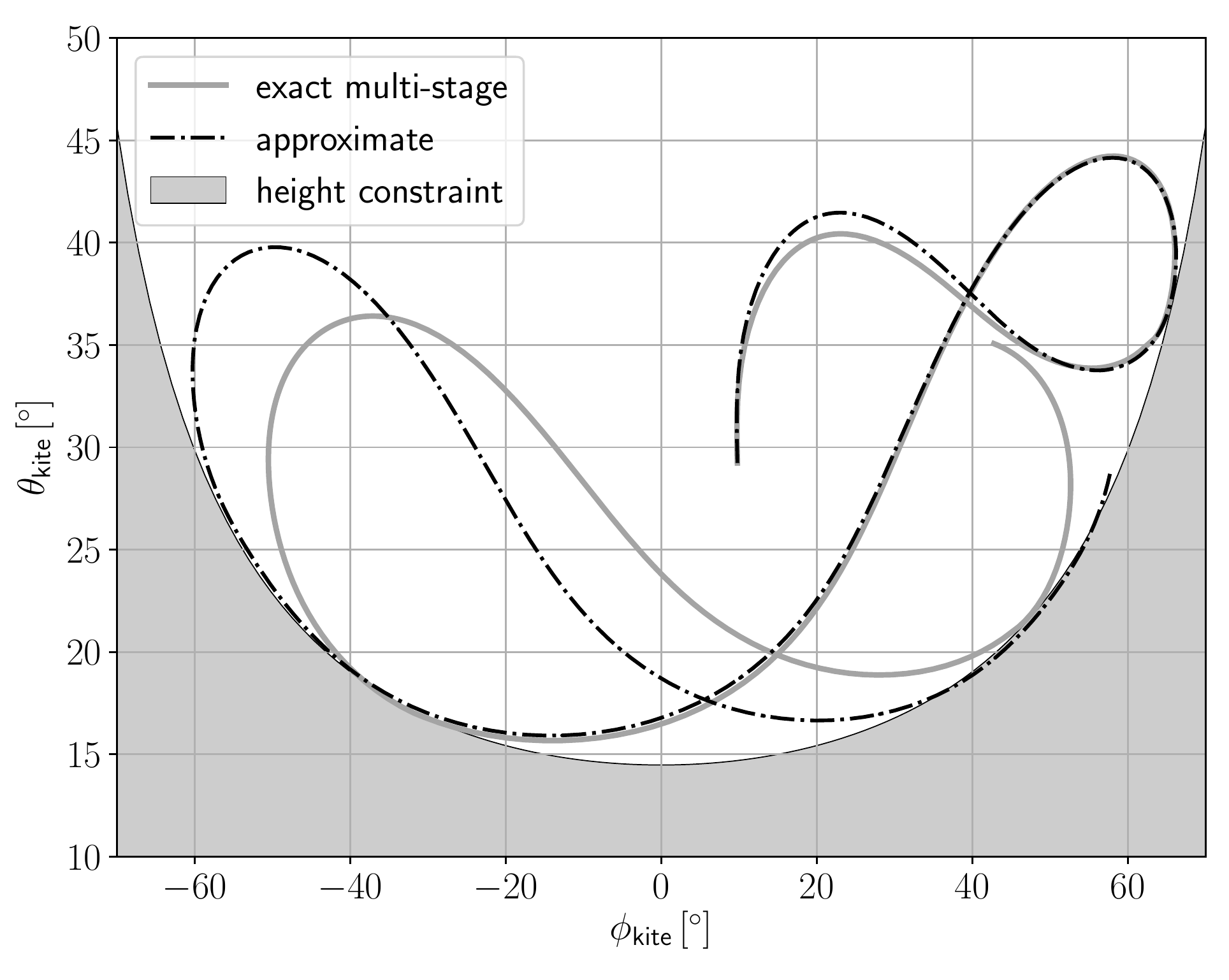}
\caption{$\eta = \SI{0}{\metre}$}
\label{fig:closed_loop_eta_0}
\end{subfigure}
\hspace*{\fill}
\begin{subfigure}{0.49\textwidth}
\includegraphics[width=.99\linewidth, trim=0.0cm 0.0cm 0.0cm 0.0cm, clip]{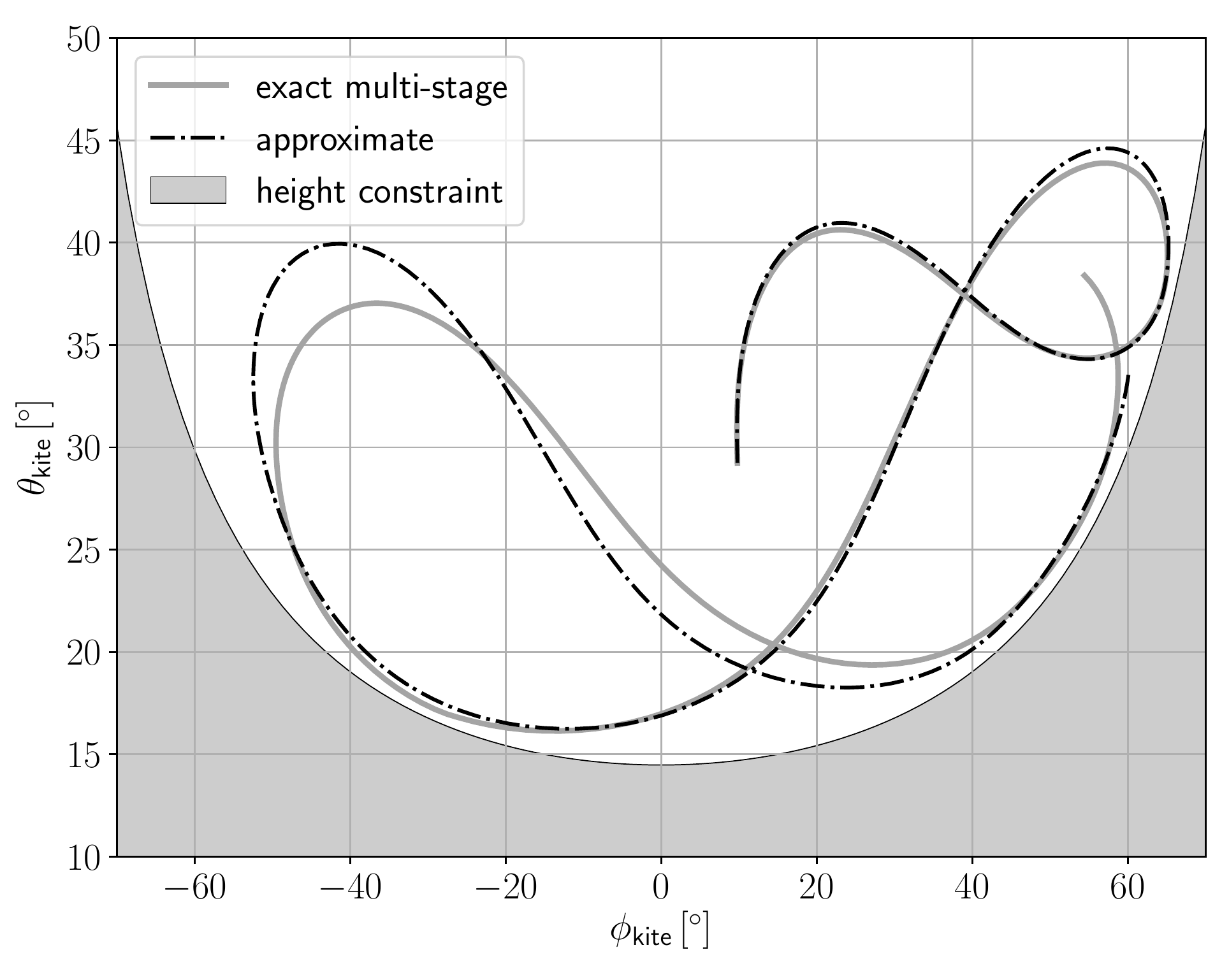}
\caption{$\eta = \SI{4}{\metre}$}
\label{fig:closed_loop_eta_4}
\end{subfigure}
\caption{Comparison of the exact multi-stage NMPC and its deep learning-based approximation for one sample $w$ and two different choices of the back-off parameter $\eta$. By choosing no backoff ($\eta=\SI{0}{\metre}$) the kite is often operated at the bound which leads to frequent constraint violations due to the estimation and uncertainty discretization errors $\epsilon_{\text{est}}$ and $\epsilon_{\text{ms}}$. The constraint violations by the neural network controller $\kappa_{\text{dnn},0}$ are more significant as it is additionally affected by the approximation error $\epsilon_{\text{approx}}$ (a). By introducing a backoff of $\eta=\SI{4}{\metre}$ the impact of the three error sources is mitigated which enables the probabilistically safe operation of the kite with both controllers (b).} \label{fig:comp_exact_approx}
\end{center}
\end{figure}

\subsection{Verification of a deep learning-based embedded output-feedback robust NMPC}
\label{subsec:verification_kite}

Because of the approximation errors, measurement and estimation errors as well as the errors derived from the multi-stage formulation, we refrain from a worst-case deterministic analysis and resort to the probabilistic verification scheme based on closed-loop trajectories presented in Section~\ref{sec:validation}.
We consider four possible values for the backoff hyper-parameter $\eta$, i.e. $\eta \in \{\SI{0}{\metre},\SI{2}{\metre},\SI{4}{\metre},\SI{6}{\metre}\}$.
This leads to a family of $M=4$ deep learning-based approximate controllers $\kappa_{\text{dnn},\eta}$.
Each one of these controllers was obtained training on data sets $\mathcal{T}_{\text{opt},\eta}$ containing 80000 data pairs each.
The resulting controllers were analyzed for $N$ i.i.d. scenarios $w^{(j)}$, $j=1,\ldots,N$ corresponding to $N$ closed-loop simulations under the dynamics presented in~\eqref{eq_kite_model}, where the uncertain parameters of the model and the initial conditions are drawn according to the distributions given in Table~\ref{tab:parameters} and first row of Table~\ref{tab:uncertainty_sampling} respectively. Since the height constraint~\eqref{eq:height_constraint} is the most critical constraint, we define the performance indicator:
\begin{align}\label{eq:performance_indicator_verification}
    \phi(w;N_{\text{sim}},\kappa_{\text{dnn},\eta}) = \max_{j=0,\dots,N_{\text{sim}}}(h_{\text{min}}- h(x(j,w))),
\end{align}
where $x(j,w)$ is the state trajectory at sampling time $j$ caused by scenario $w$ using controller $\kappa_{\text{dnn},\eta}$.
The performance indicator~\eqref{eq:performance_indicator_verification} extracts the largest violation of the minimum height $h_{\text{min}}$, if a violation occurs, or the closest value to $h_{\text{min}}$ throughout one scenario.
Each scenario has a duration of {\SI{60}{\second}} which means $N_{\text{sim}}= 400$.
To consider a controller probabilistically safe, we require that the probabilistic performance indicator satisfies:
\begin{align}\label{eq:requirement_verification}
    \text{Pr}_{\mathcal{W}}(\phi(w;N_{\text{sim}},\kappa_{\text{dnn},\eta})>0) \leq \epsilon,
\end{align}
with confidence $1-\delta$ for a randomly sampled scenario $w$ according to $\text{Pr}_\mathcal{W}$.
Following the notation of Theorem \ref{theo:probabilistic}, the performance indicators corresponding to backoff parameters $ \{\SI{0}{\metre},\SI{2}{\metre},\SI{4}{\metre},\SI{6}{\metre}\}$ are collected into vectors $\{\vv{v_1}, \vv{v_2}, \vv{v_3},\vv{v_4} \}$ respectively. We consider a value of the discarding parameter $r=4$. That is, a controller is probabilistically validated if no more than 3 simulations violate the height constraint. For these specifications ($\epsilon=0.02$, $\delta=1\times10^{-6}$, and $r=4$), $N=1388$ samples are required (see~\eqref{eq:number_samples}).
The family of controllers was evaluated for 1388 i.i.d. scenarios $w^{(j)}$ and the results are summarized in Table~\ref{tab:results_verification}.
If no backoff is considered ($\eta=\SI{0}{\metre}$) the exact multi-stage NMPC operates often at the constraint bound which leads to small violations of the height constraint as $\epsilon_{\text{ms}}$ and $\epsilon_{\text{est}}$ are ignored. The corresponding approximate controller $\kappa_{\text{dnn},0}$ is additionally affected by $\epsilon_{\text{approx}}$~\eqref{eq:errors_dnn} which leads to violations of the height constraint in more than half of the scenarios when applied. Exemplary trajectories for the exact multi-stage NMPC and the approximate controller for one scenario are visualized in Figure~\ref{fig:closed_loop_eta_0}.
By considering $\eta=\SI{2}{\metre}$, the amount of violations can be significantly reduced to 8 scenarios, which shows the importance of the backoff parameter.
However, the performance of $\kappa_{\text{dnn},2}$ is not considered probabilistically safe because after discarding the allowed number of worst-case simulation runs, we get $\psi(\mathbf{v_2},4) = \SI{0.273}{\metre} > \SI{0}{\metre}$.
With larger backoffs $\eta=\SI{4}{\metre}$ and $\eta=\SI{6}{\metre}$, we obtain two probabilistically safe controllers with performance indicator levels $\psi(\mathbf{v_3},4) = \SI{-0.316}{\metre}$ and $\psi(\mathbf{v_4},4) = \SI{-1.818}{\metre}$, respectively.
For the same scenario $w$ as in Figure~\ref{fig:closed_loop_eta_0}, the trajectories of exact multi-stage NMPC with $\eta=\SI{4}{\metre}$ and $\kappa_{\text{dnn},4}$ are depicted in Figure~\ref{fig:closed_loop_eta_4}.
Due to the backoff, the kite is keeping a safety distance to the constraint bound and the impact of $\epsilon_{\text{ms}}$ and $\epsilon_{\text{est}}$ does not directly lead to constraint violations.
Also the trajectory of the approximate controller does not violate the trajectories despite being affected by the additional approximation error $\epsilon_{\text{approx}}$.
The preferred deep learning-based controller is $\kappa_{\text{dnn},4}$ due to the higher average tether thrust $T_{\text{F}}$ provided. By introducing a performance indicator level for the average thrust per simulation run:
\begin{align}\label{eq:performance_indicator_thrust}
   \phi_{T_{\text{F}}}(w; N_{\text{sim}}, \kappa) = \frac{1}{N_{\text{sim}}}\sum_{k = 0}^{N_{\text{sim-1}}} -T_{\text{F}}(k),
\end{align}
it is possible to obtain probabilistic statements about the performance in the same fashion as for violation of the height constraint.
Using the parameters $\delta=1\times10^{-6}$, $\epsilon=0.02$, $M=4$ and $r=4$ we obtain, for the controller $\kappa_{\text{dnn},4}$, that with confidence $1-\delta$, the probability that the average thrust for a simulation run of \SI{60}{\second} duration is lower than \SI{111.346}{\kilo\newton} is not larger than $\epsilon=0.02$. 
 A smaller number of samples is required if the discarding parameter $r$ is set equal to 1. However, this leads to more conservative results because violations of the height constraint occur throughout the closed loop simulations used for verification. This is even worse when the performance index is a binary function determining if the trajectories are admissible or not. In this case, the obtained results are often not informative because in a binary setting with $r=1$, a single violated trajectory out of $N$ determines that the controller does not meet the probabilistic constraints. Larger values for $r$, along with the consideration of non-binary violation performance indexes, provide more informative results. One more advantage of the proposed probabilistic method is that a family of controllers can be evaluated in parallel in the closed loop for the same set of sampled scenarios. This can reduce the verification effort significantly, if drawing samples $w$ from $\mathcal{W}$ is costly or the closed-loop experiments have a long duration.


\begin{table}
\caption{Comparison of the members of a deep learning-based based family of controllers defined by $M=4$ different choices of the backoff parameter $\eta = \{\SI{0}{\metre},\SI{2}{\metre},\SI{4}{\metre},\SI{6}{\metre}\}$. The parameters for the probabilistic safety certificate were chosen to $\epsilon=0.02$ and $\delta=1\times10^{-6}$. The necessary number of samples for 3 discarded worst-case runs ($r=4$) is $N=1388$ and computed via~\eqref{eq:number_samples}.}\label{tab:results_verification}
\begin{center}
\begin{tabular}{c c c c c}\label{tab:results_learning}
controller & $\kappa_{\text{dnn},0}$ & $\kappa_{\text{dnn},2}$ & $\kappa_{\text{dnn},4}$ & $\kappa_{\text{dnn},6}$ \\
\midrule
feasible trajectories & 660/1388 & 1380/1388 & 1385/1388 & 1387/1388\\
$\psi(\mathbf{v},4)$ [\SI{}{\metre}] & 1.682 & 0.273 & -0.316 & -1.818\\
$T_{\text{F}}$ (avg.) [\SI{}{\kilo\newton}] & 227.516 & 225.997 & 224.185 & 222.179 \\
probabilistically safe & No & No & Yes & Yes \\
\bottomrule
\end{tabular}
\end{center}
\end{table}

\subsection{Robustness of the probabilistic validation scheme}

All obtained probabilistic guarantees are only valid if the assumptions about the probability density functions (PDFs) of $\mathcal{W}$ from which the scenarios $w$ are drawn are correct.
For the verification, the $N$ closed-loop simulations were generated using the dynamics presented in~\eqref{eq_kite_model} and the different $\kappa_{\text{dnn},\eta}$ controllers. The uncertain parameters of the model and the initial conditions were drawn according to the distributions given in Table~\ref{tab:parameters} and first row of Table~\ref{tab:uncertainty_sampling} respectively.

\begin{figure}
\begin{center}
\includegraphics[width=.9\linewidth, trim=0.0cm 0.0cm 0.0cm 1.5cm, clip]{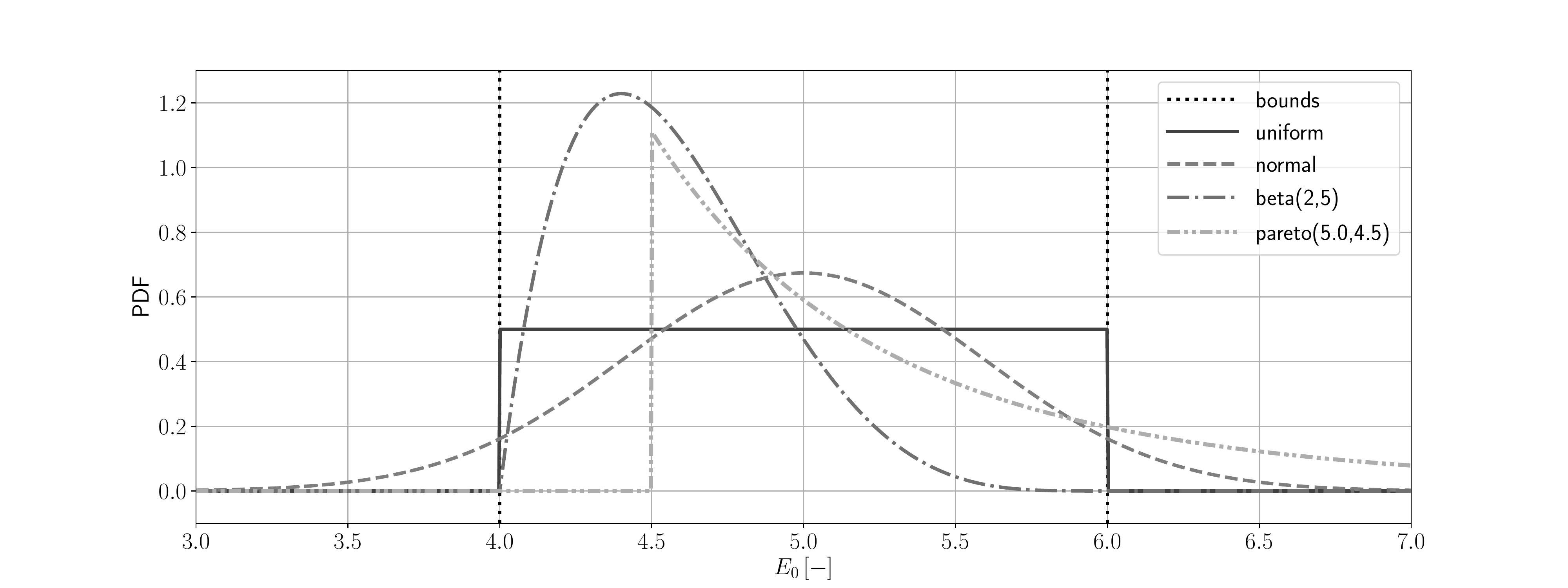}
\caption{Four different considerations of the uncertain parameter base glide ratio $E_0$ and the considered bounds in the NMPC formulation. The normal and the pareto distribution exceed the considered bounds.}
\label{fig:PDFS}
\end{center}
\end{figure}

To test the robustness of the probabilistic statements w.r.t. to wrong assumptions about the PDFs, the performance of the approximate controllers $\kappa_{\text{dnn},\eta}$ is evaluated using not the distribution of the first row of Table~\ref{tab:uncertainty_sampling}, but the second (normal distribution), the third (beta distribution) and the fourth one (pareto distribution). 
The first parameter in the description of the beta distribution is the scaling and the second parameter is the offset, e.g. $$\theta_0 = 2.0 \cdot \text{beta}(2,5) + 28.0.$$
The long-tailed pareto distribution is also described with two parameters, where the first one is the tail index and the second one is the scaling, e.g. $$\theta_0 = \text{pareto}(5.0) + 28.0.$$
The possible extreme values of samples from the space of beta distributions $\mathcal{W}_{\text{beta}}$ are identical with those when sampling from the space of uniform distributions $\mathcal{W}$, see Figure~\ref{fig:PDFS}.
In case of sampling from $\mathcal{W}_{\text{normal}}$ and $\mathcal{W}_{\text{pareto}}$, which have infinite support, the occurrence of values in $w$ which exceed the bounds of the scenarios considered in the robust MPC formulation and the verification scenarios is likely, which highlights the importance of including the discarding parameter $r$.
The four different considered PDFs including the bounds applied in the NMPC formulation are shown in Figure~\ref{fig:PDFS} for the example base glade ratio $E_0$.

The results corresponding to drawing 1388 scenarios from each of the distributions $\mathcal{W}_{\text{normal}}$, $\mathcal{W}_{\text{beta}}$ and $\mathcal{W}_{\text{pareto}}$, and evaluating the approximate controller $\kappa_{\text{dnn},4}$ are given in Table~\ref{tab:uncertainty_sampling}.
For the case of $\mathcal{W}_{\text{normal}}$ and $\mathcal{W}_{\text{beta}}$ one simulation run violates the height constraint, while three simulation runs violate the constraints for $\mathcal{W}_{\text{pareto}}$.
This means that the probabilistic requirements for the safety certificate ($\epsilon=0.02$, $\delta=1\times10^{-6}$, $M=4$, $r=4$) hold for all alternative choices of distributions.
This shows that neither the training of the network nor the verification approach fails catastrophically when the statistical assumptions are not exactly fulfilled.


\subsection{Embedded implementation}

One of the major advantages of learning the complex optimal control law via deep neural networks is the reduction of computational load and the fast evaluation.
The computation of the control input is reduced from solving an optimization problem to one matrix-vector multiplication per layer and the evaluation of the \emph{tanh}-function.
This enables the implementation of a probabilistically validated, approximate robust nonlinear model predictive control on limited hardware such as microcontrollers or field programmable gate arrays (FPGAs).
We deployed the approximate controller on a low-cost microcontroller (ARM Cortex-M3 32-bit) running with a frequency of \SI{89}{\mega\hertz} with \SI{96}{\kilo\byte} RAM.
The memory footprint of both the EKF and the neural network that describes the approximate robust NMPC is only \SI{67.0}{\kilo\byte} of the \SI{512}{\kilo\byte} flash memory.
The average time needed to evaluate the neural network was \SI{32.1}{\milli\second} (max. evaluation time: \SI{33.0}{\milli\second}) and the average evaluation time for one EKF step was \SI{28.3}{\milli\second} (max. evaluation time: \SI{30.0}{\milli\second}), which shows that the proposed controller is real-time capable with a worst-case evaluation time of \SI{63.0}{\milli\second}.
We analyzed the impact of the evaluation time on the safety by simulating the kite for the same 1388 scenarios considered in Table~\ref{tab:results_verification} drawn from the uniform distribution, but by applying the computed control inputs with a delay of $t_{\text{delay}} = \SI{65.0}{\milli\second}$, emulating a hardware-in-the-loop setting.
We rounded the time delay up to \SI{65.0}{\milli\second} to account for possible time measurement errors.
To deal with the additional error $\epsilon_{\text{delay}}$ caused by $t_{\text{delay}}$, we chose $\eta=\SI{6}{\metre}$.
Out of the 1388 simulated scenarios, 1374 were free of violations.
This means the controller violates the height constraint in $0.86\%$ of the cases, which is less than the probabilistically guarantee $\epsilon=0.02$ chosen in Section~\ref{subsec:verification_kite}, despite of the additional errors induced through the delay.
If the performance needs to be further improved for the hardware-in-the-loop setting, training data for the controller could be generated where the deterministic $t_{\text{delay}}$ is incorporated in the NMPC formulation. This is an additional advantage of the proposed approach, because the evaluation time of a given neural network is deterministic and can be known in advance. Additional measures to counteract the impact of delayed application of the control inputs such as advanced-step NMPC~\cite{zavala2009advanced} or the real-time iteration scheme~\cite{diehl2002real} could be also incorporated in the scheme.


\section{Conclusions and Future Work}\label{sec:conclusions}
The computation complexity related to  output-feedback robust NMPC controllers is prohibitive in most cases. Instead of relying on strong assumptions on error bounds and invariant sets that cannot be verified in practice, we propose a probabilistic performance validation scheme that can be used to obtain probabilistic guarantees about the closed-loop performance of approximate robust NMPC controllers based on a tree of discrete scenarios. To enable the implementation of such controllers in real-time even on limited embedded hardware, we used deep learning to approximate the proposed robust NMPC controller.

To deal with errors related to estimation, computation of approximate reachable sets based on scenarios as well as approximation of the resulting optimization problem with a neural network, we tighten the original constraints of the problem using a backoff parameter. The novel probabilistic validation framework lead to less restrictive results when compared to previous approaches because of the incorporation of a discarding parameter and the consideration of non-binary performance indicators. Moreover, the required sample complexity does not depend on the dimension of the problem. The promising results for the embedded output-feedback robust NMPC of a towing kite show the potential of the proposed approach. Future work includes the definition of robust margins based on probabilistic validation techniques as well as the learning of controllers that are parameterized, for example, with a backoff parameter.

\section*{Acknowledgments}

This work was supported  by the Agencia Estatal de Investigación (AEI)-Spain  under Grant PID2019-106212RB-C41/AEI/10.13039/501100011033.
The research leading to these results has received funding from from the Deutsche Forschungsgemeinschaft (DFG, German Research Foundation) under grant agreement number 423857295.



\subsection*{Financial disclosure}

None reported.

\subsection*{Conflict of interest}

The authors declare no potential conflict of interests.



\bibliography{refs_miqp}%

\begin{thebibliography}{10}
\providecommand{\url}[1]{#1}
\csname url@samestyle\endcsname
\providecommand{\newblock}{\relax}
\providecommand{\bibinfo}[2]{#2}
\providecommand{\BIBentrySTDinterwordspacing}{\spaceskip=0pt\relax}
\providecommand{\BIBentryALTinterwordstretchfactor}{4}
\providecommand{\BIBentryALTinterwordspacing}{\spaceskip=\fontdimen2\font plus
\BIBentryALTinterwordstretchfactor\fontdimen3\font minus
  \fontdimen4\font\relax}
\providecommand{\BIBforeignlanguage}[2]{{%
\expandafter\ifx\csname l@#1\endcsname\relax
\typeout{** WARNING: IEEEtran.bst: No hyphenation pattern has been}%
\typeout{** loaded for the language `#1'. Using the pattern for}%
\typeout{** the default language instead.}%
\else
\language=\csname l@#1\endcsname
\fi
#2}}
\providecommand{\BIBdecl}{\relax}
\BIBdecl

\bibitem{campo1987}
P.~J. Campo and M.~Morari, ``Robust model predictive control,'' in \emph{Proc.
  of the American Control Conference}, 1987, pp. 1021--1026.

\bibitem{lee1997}
J.~H. Lee and Z.~H. Yu, ``Worst-case formulations of model predictive control
  for systems with bounded parameters,'' \emph{Automatica}, vol.~33, no.~5, pp.
  763--781, 1997.

\bibitem{mayne2005}
D.~Q. Mayne, M.~M. Seron, and S.~V. Rakovic, ``Robust model predictive control
  of constrained linear systems with bounded disturbances,'' \emph{Automatica},
  vol.~41, pp. 219 -- 224, 2005.

\bibitem{rakovic2011}
S.~Rakovic, B.~Kouvaritakis, M.~Cannon, C.~Panos, and R.~Findeisen, ``Fully
  parameterized tube {MPC},'' in \emph{Proc. of the 18th IFAC World Congress
  Milano}, 2011, pp. 197--202.

\bibitem{Fleming2015}
J.~Fleming, B.~Kouvaritakis, and M.~Cannon, ``Robust tube {MPC} for linear
  systems with multiplicative uncertainty,'' \emph{IEEE Transactions on
  Automatic Control}, vol.~60, no.~4, pp. 1087--1092, 2015.

\bibitem{scokaert1998}
P.~Scokaert and D.~Mayne, ``Min-max feedback model predictive control for
  constrained linear systems,'' \emph{IEEE Transactions on Automatic Control},
  vol.~43, no.~8, pp. 1136--1142, 1998.

\bibitem{delaPena2005b}
D.~Mu\~noz de~la Pe\~na, A.~Bemporad, and T.~Alamo, ``Stochastic programming
  applied to model predictive control,'' in \emph{Proc. of the 44th IEEE
  Conference on Decision and Control}, 2005, pp. 1361--1366.

\bibitem{bernardini2009}
D.~Bernardini and A.~Bemporad, ``Scenario-based model predictive control of
  stochastic constrained linear systems,'' in \emph{Proc. of the 48th IEEE
  Conference on Decision and Control}, 2009, pp. 6333--6338.

\bibitem{lucia2013}
S.~Lucia, T.~Finkler, and S.~Engell, ``Multi-stage nonlinear model predictive
  control applied to a semi-batch polymerization reactor under uncertainty,''
  \emph{Journal of Process Control}, vol.~23, pp. 1306--1319, 2013.

\bibitem{lucia2014e}
S.~Lucia, J.~Andersson, H.~Brandt, M.~Diehl, and S.~Engell, ``Handling
  uncertainty in economic nonlinear model predictive control: a comparative
  case-study,'' \emph{Journal of Process Control}, vol.~24, pp. 1247--1259,
  2014.

\bibitem{goulart2006}
P.~Goulart, E.~C. Kerrigan, and J.~M. Maciejowski, ``Optimization over state
  feedback policies for robust control with constraints,'' \emph{Automatica},
  vol.~42, pp. 523--533, 2006.

\bibitem{LuciaPhd}
S.~Lucia, \emph{Robust Multi-stage Nonlinear Model Predictive Control}.\hskip
  1em plus 0.5em minus 0.4em\relax Shaker, 2014.

\bibitem{lucia2020stability}
S.~Lucia, S.~Subramanian, D.~Limon, and S.~Engell, ``Stability properties of
  multi-stage nonlinear model predictive control,'' \emph{Systems \& Control
  Letters}, vol. 143, p. 104743, 2020.

\bibitem{houska2011}
B.~Houska, H.~Ferreau, and M.~Diehl, ``An auto-generated real-time iteration
  algorithm for nonlinear {MPC} in the microsecond range,'' \emph{Automatica},
  vol.~47, pp. 2279--2285, 2011.

\bibitem{mattingley2012cvxgen}
J.~Mattingley and S.~Boyd, ``Cvxgen: A code generator for embedded convex
  optimization,'' \emph{Optimization and Engineering}, vol.~13, no.~1, pp.
  1--27, 2012.

\bibitem{zometa2013}
P.~Zometa, M.~K\"ogel, and R.~Findeisen, ``{muAO-MPC}: A free code generation
  tool for embedded real-time linear model predictive control,'' in \emph{Proc.
  of the American Control Conference}, 2013, pp. 5320--5325.

\bibitem{lucia2018_TII}
S.~Lucia, D.~Navarro, O.~Lucia, P.~Zometa, and R.~Findeisen, ``Optimized {FPGA}
  implementation of model predictive control using high level synthesis
  tools,'' \emph{IEEE Transactions on Industrial Informatics}, vol.~14, no.~1,
  pp. 137--145, 2018.

\bibitem{johansen2004}
T.~A. Johansen, ``{Approximate explicit receding horizon control of constrained
  nonlinear systems},'' \emph{Automatica}, vol.~40, no.~2, pp. 293--300, 2004.

\bibitem{johansen2002multi}
------, ``On multi-parametric nonlinear programming and explicit nonlinear
  model predictive control,'' in \emph{IEEE Conference on Decision and
  Control}, vol.~3, 2002, pp. 2768--2773.

\bibitem{bemporad2002}
A.~Bemporad, M.~Morari, V.~Dua, and E.~N. Pistikopoulos, ``The explicit linear
  quadratic regulator for constrained systems,'' \emph{Automatica}, vol.~38,
  no.~1, pp. 3 -- 20, 2002.

\bibitem{parisini1995}
T.~Parisini and R.~Zoppoli, ``A receding-horizon regulator for nonlinear
  systems and a neural approximation,'' \emph{Automatica}, vol.~31, no.~10, pp.
  1443--1451, 1995.

\bibitem{karg2018efficient}
B.~Karg and S.~Lucia, ``Efficient representation and approximation of model
  predictive control laws via deep learning,'' \emph{IEEE Transactions on
  Cybernetics}, vol.~50, no.~9, pp. 3866--3878, 2020.

\bibitem{Chen2018}
S.~Chen, K.~Saulnier, and N.~Atanasov, ``Approximating explicit model
  predictive control using constrained neural networks,'' in \emph{Proc. of the
  American Control Conference}, 2018, pp. 1520--1527.

\bibitem{safran2017}
I.~Safran and O.~Shamir, ``Depth-width tradeoffs in approximating natural
  functions with neural networks,'' in \emph{Proceedings of the 34th
  International Conference on Machine Learning}, 2017, pp. 2979--2987.

\bibitem{Vidyasagar97}
M.~Vidyasagar, \emph{A Theory of Learning and Generalization}.\hskip 1em plus
  0.5em minus 0.4em\relax London: Springer, 1997.

\bibitem{Tempo13}
R.~Tempo, G.~Calafiore, and F.~Dabbene, \emph{Randomized Algorithms for
  Analysis and Control of Uncertain Systems, with Applications}, 2nd~ed.\hskip
  1em plus 0.5em minus 0.4em\relax London: Springer-Verlag, 2013.

\bibitem{Grammatico:16}
S.~Grammatico, X.~Zhang, K.~Margellos, P.~Goulart, and J.~Lygeros, ``A scenario
  approach for non-convex control design,'' \emph{IEEE Transactions on
  Automatic Control}, vol.~61, no.~2, pp. 334--345, 2016.

\bibitem{Lorenzen:17}
M.~Lorenzen, F.~Dabbene, R.~Tempo, and F.~Allg\"{o}wer, ``Stochastic {MPC} with
  offline uncertainty sampling,'' \emph{Automatica}, vol.~81, pp. 176--183,
  2017.

\bibitem{Mammarella:18:Control:Systems:Technology}
M.~{Mammarella}, M.~{Lorenzen}, E.~{Capello}, H.~{Park}, F.~{Dabbene},
  G.~{Guglieri}, M.~{Romano}, and F.~{Allg\"ower}, ``An offline-sampling {SMPC}
  framework with application to autonomous space maneuvers,'' \emph{IEEE
  Transactions on Control Systems Technology}, pp. 1--15, 2018.

\bibitem{Calafiore06}
G.~Calafiore and M.~Campi, ``The scenario approach to robust control design,''
  \emph{IEEE Transactions on Automatic Control}, vol.~51, no.~5, pp. 742--753,
  2006.

\bibitem{fagiano2010}
L.~Fagiano, M.~Milanese, V.~Razza, and I.~Gerlero, ``Control of power kites for
  naval propulsion,'' in \emph{Proc. of the American Control Conference}, 2010,
  pp. 4325--4330.

\bibitem{deori2017trading}
L.~Deori, S.~Garatti, and M.~Prandini, ``Trading performance for state
  constraint feasibility in stochastic constrained control: A randomized
  approach,'' \emph{Journal of the Franklin Institute}, vol. 354, no.~1, pp.
  501--529, 2017.

\bibitem{Margellos:14}
K.~Margellos, P.~Goulart, and J.~Lygeros, ``On the road between robust
  optimization and the scenario approach for chance constrained optimization
  problems,'' \emph{IEEE Transactions on Automatic Control}, vol.~59, no.~8,
  pp. 2258--2263, 2014.

\bibitem{Alamo09}
T.~Alamo, R.~Tempo, and E.~Camacho, ``Randomized strategies for probabilistic
  solutions of uncertain feasibility and optimization problems,'' \emph{IEEE
  Transactions on Automatic Control}, vol.~54, no.~11, pp. 2545--2559, 2009.

\bibitem{TeBaDa:97}
R.~Tempo, E.~Bai, and F.~Dabbene, ``Probabilistic robustness analysis: explicit
  bounds for the minimum number of samples,'' \emph{Systems {\&} Control
  Letters}, vol.~30, pp. 237--242, 1997.

\bibitem{Alamo:15}
T.~Alamo, R.~Tempo, A.~Luque, and D.~Ramirez, ``Randomized methods for design
  of uncertain systems: sample complexity and sequential algorithms,''
  \emph{Automatica}, vol.~52, pp. 160--172, 2015.

\bibitem{alamir2018feedback}
M.~Alamir, M.~Fiacchini, I.~Queinnec, S.~Tarbouriech, and M.~Mazerolles,
  ``Feedback law with probabilistic certification for {P}ropofol-based control
  of {BIS} during anesthesia,'' \emph{International Journal of Robust and
  Nonlinear Control}, vol.~28, no.~18, pp. 6254--6266, 2018.

\bibitem{Alamir:15}
M.~Alamir, ``On probabilistic certification of combined cancer therapies using
  strongly uncertain models,'' \emph{Journal of Theoretical Biology}, vol. 384,
  pp. 59--69, 2015.

\bibitem{hertneck2018learning}
M.~Hertneck, J.~K{\"o}hler, S.~Trimpe, and F.~Allg{\"o}wer, ``Learning an
  approximate model predictive controller with guarantees,'' \emph{IEEE Control
  Systems Letters}, vol.~2, no.~3, pp. 543--548, 2018.

\bibitem{zhang2019safe}
X.~Zhang, M.~Bujarbaruah, and F.~Borrelli, ``Safe and near-optimal policy
  learning for model predictive control using primal-dual neural networks,''
  \emph{arXiv preprint arXiv:1906.08257}, 2019.

\bibitem{karg2019ecc}
B.~Karg and S.~Lucia, ``Learning-based approximation of robust nonlinear
  predictive control with state estimation applied to a towing kite,'' in
  \emph{Proc. of the European Control Conference}, 2019, pp. 16--22.

\bibitem{blondel2000survey}
V.~Blondel and J.~Tsitsiklis, ``A survey of computational complexity results in
  systems and control,'' \emph{Automatica}, vol.~36, no.~9, pp. 1249--1274,
  2000.

\bibitem{Hoeffding:63}
W.~Hoeffding, ``Probability inequalities for sums of bounded random
  variables,'' \emph{Journal of the American Statistical Association}, vol.~58,
  no. 301, pp. 13--30, 1963.

\bibitem{Alamo:18}
T.~Alamo, J.~Manzano, and E.~Camacho, ``Robust design through probabilistic
  maximization,'' in \emph{{Uncertainty in Complex Networked Systems. In Honor
  of Roberto Tempo}}, T.~Basar, Ed.\hskip 1em plus 0.5em minus 0.4em\relax
  Birkh\"auser, 2018, pp. 247--274.

\bibitem{Ahsanullah:13}
M.~Ahsanullah, V.~Nevzorov, and M.~Shakil, \emph{An introduction to Order
  Statistics}.\hskip 1em plus 0.5em minus 0.4em\relax Paris: Atlantis Press,
  2013.

\bibitem{Arnold:92}
B.~Arnold, N.~Balakrishnan, and H.~Nagaraja, \emph{A First Course in Order
  Statistics}.\hskip 1em plus 0.5em minus 0.4em\relax New York: John Wiley and
  Sons, 1992.

\bibitem{oishi2007polynomial}
Y.~Oishi, ``Polynomial-time algorithms for probabilistic solutions of
  parameter-dependent linear matrix inequalities,'' \emph{Automatica}, vol.~43,
  no.~3, pp. 538--545, 2007.

\bibitem{calafiore2011research}
G.~C. Calafiore, F.~Dabbene, and R.~Tempo, ``Research on probabilistic methods
  for control system design,'' \emph{Automatica}, vol.~47, no.~7, pp.
  1279--1293, 2011.

\bibitem{alamo2015randomized}
T.~Alamo, R.~Tempo, A.~Luque, and D.~R. Ramirez, ``Randomized methods for
  design of uncertain systems: Sample complexity and sequential algorithms,''
  \emph{Automatica}, vol.~52, pp. 160--172, 2015.

\bibitem{Alamo10ACC}
T.~Alamo, R.~Tempo, and A.~Luque, ``On the sample complexity of randomized
  approaches to the analysis and design under uncertainty,'' in
  \emph{Proceedings of the American Control Conference}, Baltimore, USA, 2010.

\bibitem{rawlings2009}
J.~Rawlings and D.~Mayne, \emph{Model predictive control: Theory and
  design}.\hskip 1em plus 0.5em minus 0.4em\relax Nob Hill Pub., 2009.

\bibitem{althoff2014}
M.~Althoff and B.~Krogh, ``Reachability analysis of nonlinear
  differential-algebraic systems,'' \emph{IEEE Transactions on Automatic
  Control}, vol.~59, pp. 371--383, 2014.

\bibitem{sahlodin11}
A.~Sahlodin and B.~Chachuat, ``Convex/concave relaxations of parametric {ODE}s
  using {T}aylor models,'' \emph{Computers \& Chemical Engineering}, vol.~35,
  no.~5, pp. 844 -- 857, 2011.

\bibitem{hewing2018cdc}
L.~{Hewing} and M.~N. {Zeilinger}, ``Stochastic model predictive control for
  linear systems using probabilistic reachable sets,'' in \emph{2018 IEEE
  Conference on Decision and Control (CDC)}, 2018, pp. 5182--5188.

\bibitem{subramanian2018synergistic}
S.~Subramanian, S.~Lucia, and S.~Engell, ``A synergistic approach to robust
  output feedback control: Tube-based multi-stage {NMPC},''
  \emph{IFAC-PapersOnLine}, vol.~51, no.~18, pp. 500--505, 2018.

\bibitem{barron1993universal}
A.~R. Barron, ``Universal approximation bounds for superpositions of a
  sigmoidal function,'' \emph{IEEE Transactions on Information theory},
  vol.~39, no.~3, pp. 930--945, 1993.

\bibitem{houska2006}
B.~Houska and M.~Diehl, ``Optimal control of towing kites,'' in \emph{Proc. of
  the 45th IEEE Conference on Decision and Control}, 2006, pp. 2693--2697.

\bibitem{erhard2013}
M.~Erhard and H.~Strauch, ``Control of towing kites for seagoing vessels,''
  \emph{IEEE Transactions on Control Systems Technology}, vol.~21, pp.
  1629--1640, 2013.

\bibitem{fagiano2014automatic}
L.~Fagiano and C.~Novara, ``Automatic crosswind flight of tethered wings for
  airborne wind energy: a direct data-driven approach,'' \emph{IFAC Proceedings
  Volumes}, vol.~47, no.~3, pp. 4927--4932, 2014.

\bibitem{costello2013}
S.~Costello, G.~Fran\c{c}ois, and D.~Bonvin, ``Real-time optimization for
  kites,'' in \emph{Proc of the IFAC International Workshop on Periodic Control
  Systems (PSYCO)}, 2013, pp. 64--69.

\bibitem{costello2017crosswind}
S.~Costello, G.~Fran{\c{c}}ois, and D.~Bonvin, ``Crosswind kite control--a
  benchmark problem for advanced control and dynamic optimization,''
  \emph{European Journal of Control}, vol.~35, pp. 1--10, 2017.

\bibitem{chollet2015keras}
F.~Chollet \emph{et~al.}, ``Keras,'' \url{https://github.com/fchollet/keras},
  2015.

\bibitem{tensorflow2015-whitepaper}
\BIBentryALTinterwordspacing
M.~A. et~al., ``{TensorFlow}: Large-scale machine learning on heterogeneous
  systems,'' 2015, software available from tensorflow.org. [Online]. Available:
  \url{http://tensorflow.org/}
\BIBentrySTDinterwordspacing

\bibitem{kingma2014}
D.~P. Kingma and J.~Ba, ``Adam: {A} method for stochastic optimization,'' in
  \emph{{Proceedings of the 3rd International Conference on Learning
  Representations (ICLR)}}, 2014.

\bibitem{glorot2010understanding}
X.~Glorot and Y.~Bengio, ``Understanding the difficulty of training deep
  feedforward neural networks,'' in \emph{Proceedings of the thirteenth
  international conference on artificial intelligence and statistics}, 2010,
  pp. 249--256.

\bibitem{kandasamy2018neural}
K.~Kandasamy, W.~Neiswanger, J.~Schneider, B.~Poczos, and E.~P. Xing, ``Neural
  architecture search with bayesian optimisation and optimal transport,'' in
  \emph{Advances in neural information processing systems}, 2018, pp.
  2016--2025.

\bibitem{zavala2009advanced}
V.~M. Zavala and L.~T. Biegler, ``The advanced-step nmpc controller:
  Optimality, stability and robustness,'' \emph{Automatica}, vol.~45, no.~1,
  pp. 86--93, 2009.

\bibitem{diehl2002real}
M.~Diehl, H.~G. Bock, J.~P. Schl{\"o}der, R.~Findeisen, Z.~Nagy, and
  F.~Allg{\"o}wer, ``Real-time optimization and nonlinear model predictive
  control of processes governed by differential-algebraic equations,''
  \emph{Journal of Process Control}, vol.~12, no.~4, pp. 577--585, 2002.

\end{thebibliography}


\section*{Author Biography}

\begin{biography}{\includegraphics[width=68pt]{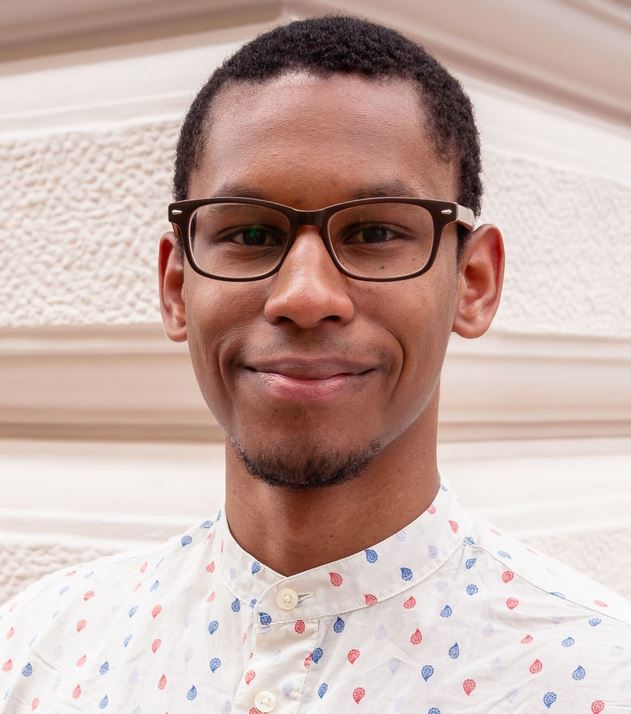}}{\textbf{Benjamin Karg.} was born in Burglengenfeld, Germany, in 1992.
He received the B.Eng. degree in mechanical engineering from Ostbayerische Technische Hochschule Regensburg, Regensburg, Bavaria, Germany, in 2015, and his M.Sc. degree in systems engineering and engineering cybernetics from Otto-von-Guericke Universit\"at, Magdeburg, Saxony-Anhalt, Germany, in 2017.
He is working as a research assistant, formerly at the laboratory "Internet of Things for Smart Buildings", Technische Universit\"at Berlin, Germany, and currently at the chair "Process Automation Systems" at Technische Universit\"at Dortmund, Dortmund, Germany, where he pursues his PhD.
He is also member of the Einstein Center for Digital Future.

His research is focused on control engineering, artificial intelligence and edge computing for IoT-enabled cyber-physical systems.}
\end{biography}

\begin{biography}{\includegraphics[width=68pt]{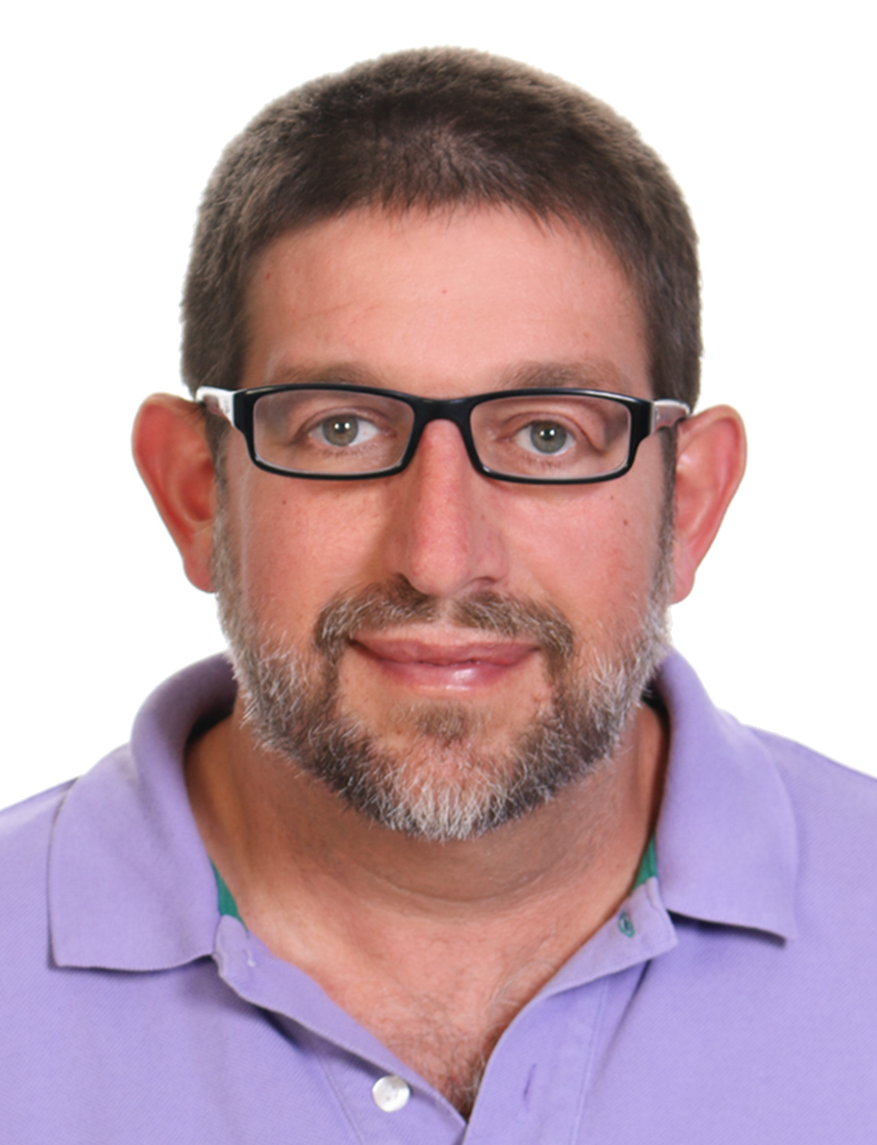}}{\textbf{Teodoro Alamo} was born in Spain in 1968. He received the M.Eng. degree in telecommunications engineering from the Polytechnic University of Madrid, Spain, in 1993 and the Ph.D. degree in telecommunications engineering from the University of Seville, Spain, in 1998.
From 1993 to 2000, he was an Assistant Professor, an Associate Professor from 2001 to 2010, and has been a Full Professor since March 2010 with the Department
of System Engineering and Automatic Control, University of Seville. He was at the Ecole Nationale Superiore des Telecommunications (Telecom Paris) from September 1991 to May 1993. Part of his Ph.D. was done at RWTH Aachen, Germany, from June to September 1995. 

He is the author or coauthor of more than 200 publications including books, book chapters, journal papers, conference proceedings, and educational books. (google scholar profile available at http://scholar.google.es/citations?user=W3ZDTkIAAAAJ\&hl=en). He has co-funded the spin-off company Optimal Performance (University of Seville, Spain). His current research interests include decision making, model predictive control, machine learning, randomized algorithms, and optimization strategies.}
\end{biography}

\begin{biography}{\includegraphics[width=68pt]{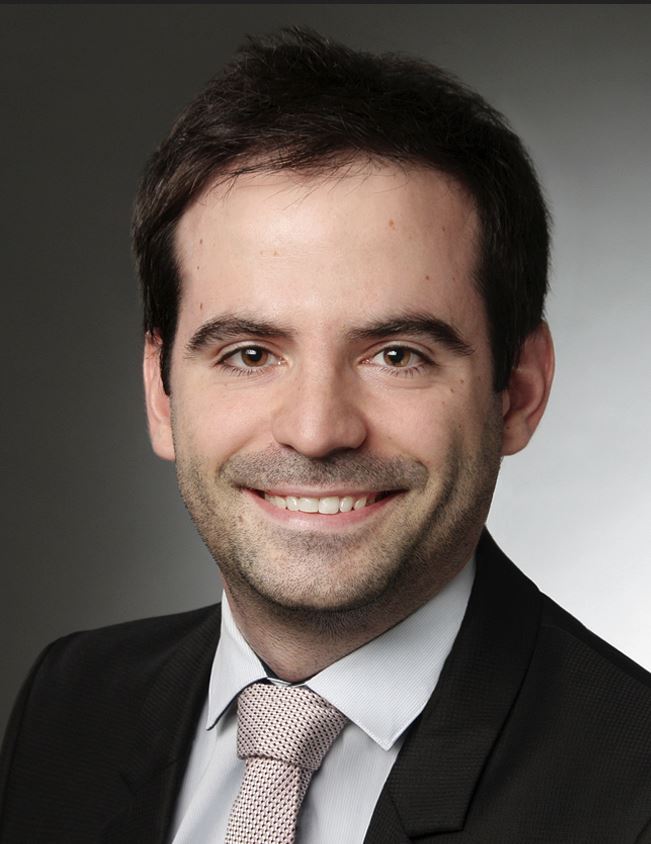}}{\textbf{Sergio Lucia.} received the M.Sc. degree
in electrical engineering from the University of
Zaragoza, Zaragoza, Spain, in 2010, and the
Dr. Ing. degree in optimization and automatic
control from the Technical University of Dortmund,
Dortmund, Germany, in 2014.
He joined the Otto-von-Guericke Universit\"at Magdeburg and visited the Massachusetts
Institute of Technology as a Postdoctoral Fellow.
In 2017, he was appointed Assistant Professor
at the Technische Universit\"at Berlin, Germany.
Since October 2020, he is a Professor and head of the Laboratory of Process Automation Systems at the Technische Universit\"at Dortmund, Germany.
His research interests include decision-making under uncertainty, distributed control, as well as the interplay between machine learning techniques and control theory.
Dr. Lucia is currently Associate Editor of the Journal of Process Control.}
\end{biography}
\end{document}